\documentclass[submission, Phys]{SciPost}
\usepackage{physics}

\begin{document}

\begin{center}{\Large \textbf{
Kondo breakdown as an entanglement transition driven by continuous measurement
}}\end{center}

\begin{center}
Debraj Debata\textsuperscript{1},
Abhirup Mukherjee\textsuperscript{1},
Siddhartha Lal\textsuperscript{1*}
\end{center}

\begin{center}
{\bf 1} Department of Physical Sciences, Indian Institute of Science Education and Research Kolkata, Mohanpur Campus, West Bengal- 741246, India
\\
* slal@iiserkol.ac.in
\end{center}

\begin{center}
\today
\end{center}


\section*{Abstract}
{\bf
We study the breakdown of Kondo screening by a local magnetic field from the perspective of a measurement-driven entanglement transition in a monitored quantum system. Here, the Kondo coupling leads to the growth in entanglement of an impurity spin with it's fermionic environment, while the local field plays the role of a continuous observer. Using a non-perturbative Unitary Renormalization Group (URG) approach, we derive coupled renormalization-group flow equations for the Kondo exchange and the local field, and obtain a field-dependent RG phase diagram. The RG flows separate a low-energy Kondo-screened phase, where the impurity is absorbed into the Fermi sea and forms an entangled singlet with the conduction bath, from a polarized local-moment phase in which screening is frustrated and impurity-bath entanglement is suppressed. We identify the fixed-point Hamiltonians governing the two phases and the critical regime, and relate the transition to the emergence of a novel non-Fermi liquid. Various impurity signatures such as the spectral function and thermalisation of impurity observables are used to characterise this entanglement transition. These results offer insight into the interplay of decoherence and measurement in governing the dynamics of a prototypical quantum system.
}

\vspace{10pt}
\noindent\rule{\textwidth}{1pt}
\tableofcontents\thispagestyle{fancy}
\noindent\rule{\textwidth}{1pt}
\vspace{10pt}

\section{Introduction}
\label{sec:intro}
Measurement-induced entanglement phase transitions in monitored quantum systems~\cite{fisher2023random, potter2022entanglement, skinner2023introduction, skinner2019measurement, chan2019unitary, koh2023measurement, feng2023measurement, gullans2020dynamical, nahum2021measurement, buchhold2022revealing, poboiko2024measurement, wiersema2023measurement,tang2020measurement, nahum2023renormalization,guo2025} have gained significant attention in recent years. These transitions result from the competition between entanglement growth and continuous measurements in quantum circuits, and are typically studied through time evolution. In this setting, unitary operators drive the growth of entanglement, while projective measurement operators disrupt the growth as time progresses. The competition leads to a phase transition that depends on the rate at which the measurements are conducted. The entanglement encoded within the state of the system remains robust until the measurement frequency exceeds a critical threshold, beyond which the system transitions into a pure disentangled state (the quantum Zeno phase~\cite{li2018,segal2007,misrasudarshan1977}). The essence of such a transition is {\it projection by emergence}.

We show here from a renormalisation group (RG) based perspective that a similar phenomenon can be studied within the context of the frustration of Kondo screening due to the influence of an external magnetic field. The Kondo effect~\cite{kondo1964resistance} describes the effects of a magnetic exchange coupling between a local spin magnetic moment/qubit (the ``impurity") and a bath of non-interacting conduction (the environment). The ground state of the system undergoes a crossover between a fully screened $SU(2)$ spin symmetry-preserved singlet ground state at energies significantly below the Kondo temperature $T_K$, and a free spin well above it (which is susceptible to polarisation by even an infinitesimal external magnetic field) (see Ref.\cite{hewson1993} and references therein). A striking observation of the effects of an magnetic field on the Kondo effect is the splitting of the spectral density of the spin impurity as the field is tuned. First demonstrated theoretically in Ref.\cite{costi2000} as the external field exceeds approximately $0.5 T_K$, similar observations have been reported subsequently in several theoretical~\cite{filippone2018magnetic, moore2000anomalous,minamitani2012symmetry, 
zhang2018kondo,niu2015kondo, ovchinnikov1999strong, kretinin2011spin, rosch2003spectral,mora2015fermi} and even a few recent experimental studies~\cite{zhang2013temperature,bagchiPRL2024}. 

As we will show here, this system exhibits an entanglement phase transition when the Kondo screening mechanism undergoes a breakdown upon the application of an external magnetic field on the impurity. In this system, the emergence of Kondo screening is responsible for the growth in entanglement of the local moment with the conduction bath environment. On the other hand, the external field acts analogously to the measurement rate in tuning towards the entanglement transition from a metallic entangled state to a polarized local moment state. This quantum phase transition is also a simple example of fermionic criticality: the Fermi volume of the conduction bath (a topological quantum number) increases discontinuously across the transition upon going from from decoupled local moment ground state to an entangled singlet, as the impurity spin is absorbed within the Fermi volume in the latter case~\cite{martin_1982,oshikawa2000topological}. Unlike the time-evolution scenarios in quantum circuits, the measurement-induced transition in our case is studied directly from the RG evolution of the system's Hamiltonian. Related RG approaches on measurement-driven phase transitions in monitored quantum systems can be found in Refs.~\cite{tang2020measurement} (1D Bose-Hubbard model), \cite{nahum2023renormalization} (a matrix model based on the physics of spin glasses), \cite{guo2025} (interacting fermions in 1D) and \cite{nahum2023renormalization} (random tensor network models). There is, however, an important difference worth noting. In these works, the outcome of an intermittent measurement process is treated as random in nature, rendering formal similarities to the physics of disordered systems. In our case, however, we study the effects of a continuous measurement whose outcome is deterministic. 
\par
The frustrated Kondo impurity problem at hand also offers a novel platform by which to understand how the competing influences of measurement by a continuous observer (the magnetic field) and decoherence~\cite{schlosshauer2019quantum, bacciagaluppi-decoherence, mazzola2010sudden, van2005intrinsic, van2006relation} due to entanglement with an environment (the electronic conduction bath) shape the dynamics of a quantum system (the impurity spin). An outstanding challenge in the foundations of quantum mechanics involves understanding how a quantum system turn classical due to the influence of it's environment. There is a venerable history of using simple models to meet this challenge, e.g., the spin-boson problem (see Ref.\cite{leggett1987dynamics} and references therein) models the influence of friction arising from a bosonic bath on the tunneling dynamics of a two-level quantum system. Indeed, it was established that the spin-boson model can be mapped onto the spin-1/2 single channel Kondo problem~\cite{guinea1985bosonization}: the dissipative coupling and tunnel amplitude of the spin-boson model can be mapped onto the Ising and spin-flip couplings of the Kondo model. A renormalisation group calculation ~\cite{caldeira1983quantum,guinea1985bosonization} reveals that a quantum phase transition separates the dissipation dominated regime of the spin boson model (corresponding to a ferromagnetic coupling in the Kondo model) where the spin decouples from the bosonic bath and forms a free local magnetic moment from a tunneling dominated regime (corresponding to an  antiferromagnetic coupling in the Kondo model) where the spin forms a maximally entangled quantum ground state together with the bath. Summarised eloquently by Mermin~\cite{mermin1991}, this serves as a prototypical demonstration of how a dissipative coupling from a bath can quench the quantum dynamics of a simple quantum system.

In a similar vein, we seek to understand whether the quantum dynamics of the spin arising from an antiferromagnetic Kondo coupling with its fermionic environment (i.e., the tunnel dominated regime of the spin-boson model) is separated from a magnetic-field dominated regime (in which all spin dynamics are quenched) by a measurement-induced quantum phase transition. Following Ref.\cite{duruisseau2023}, it is evident that the real-space local $SU(2)$ invariant two-body Kondo interaction between the impurity and its environment does not possess the global tensor-product structure necessary for the existence of classical pointer (up and down) states that are robust under the effects of decoherence~\cite{zurekRMP2003}. Starting from an initial decoupled state of the spin (the ground state of the quantum Zeno phase~\cite{segal2007}), a RG relevant antiferromagnetic Kondo interaction will lead to a maximally entangled singlet ground state, and the pointer states of the impurity spin can only exist when the local field causes Kondo breakdown. Upon tracing out the fermionic bath, this evolution will be observed in the reduced density matrix of the spin as thermalization towards a maximally mixed quantum mixture~\cite{rigol2009breakdown,wang2024eigenstate,
nandkishore2015many,deutsch2018eigenstate,kaufman2016quantum,abanin2019colloquium,weymann2015thermalization} due to the effects of decoherence. In this way, our study (of even a prototypical system) is relevant to present day noisy intermediate-scale quantum computation involving interacting qubits placed in such a decohering environment, and whose outputs must be measured during the readout process~\cite{BhartiRMP2022}.

Here, we meet these challenges by employing the non-perturbative Unitary Renormalization Group (URG) method developed recently by some of us~\cite{mukherjee2020holographic-A, mukherjee2020holographic-B}. This method has previously been applied to various models of strongly correlated electrons and quantum spins (including the single channel Kondo model)~\cite{mukherjee2020scaling-A, mukherjee2020scaling-B, pal2019correlated, 
patra2021origin, mukherjee2021fermionic, mukherjee2022unveiling, mukherjee2023kondo}. We now lay out the structure of the work. We first introduce the model and the URG method briefly in Section \ref{introduction}. In Section \ref{kondobreakdownsection}, we present the major results obtained from the URG analysis (e.g., RG equations, RG phase diagram and fixed point Hamiltonians of various phases). We then present the effective theory for the non-Fermi liquid gapless excitations at the quantum critical point in Section \ref{criticalpointsection}. In Section \ref{spectraltherm}, we investigate the transition from the perspective of various spectral features (impurity spectral function, quasiparticle residue) as well as the breakdown in eigenstate thermalisation. We then employ the momentum space entanglement renormalisation group (MERG) method developed by us~\cite{mukherjee2022unveiling,PatraEntanglement2023} in Section \ref{mergsection} to present results on the transition when viewed via the RG evolution of various local observables related to the impurity spin moment and its connection to its environment. We conclude in Section \ref{conclusionssection}. Details of several calculations are presented in various appendices. Given the variety of analyses presented below, we first summarise our major results briefly

\subsection*{Summary of major results}
\begin{itemize}    
\item From a $T=0$ URG analysis, we establish rigorously that a local magnetic field can frustrate and eventually break down Kondo screening. We interpret this as a measurement-driven entanglement transition, with the field playing the role of a continuous measurement that suppresses spin-flip fluctuations of the impurity moment mediated by the bath at low temperatures and prevents the growth of impurity–bath entanglement.

\item  Observed through the computation of an effective Hamiltonian, the transition involves the breakdown of Landau quasiparticle excitations of a local Fermi liquid, and the appearance of the gapless excitations of a novel non-Fermi liquid arising from Kondo breakdown. 

\item We connect the transition to experimentally relevant features such as magnetic-field-induced splitting of the impurity spectral function and variation of the quasiparticle residue across the transition.

\item The transition is also analyzed via eigenstate thermalization-related diagnostics and the evolution of impurity observables along the renormalisation group flow. We link such diagnostics to decoherence of the impurity spin and changes in entanglement patterns between the impurity and the conduction bath.
\end{itemize}

\section{Model and URG Method}\label{introduction}
\subsection{Kondo model with local magnetic field}
The celeberated Kondo model \cite{kondo1964resistance, anderson1969exact, anderson1970exact, anderson2018poor, bulla2008numerical, wilson1975renormalization, mukherjee2022unveiling, hewson1997kondo} describes a quantum spin-$1/2$ impurity $S_d$ coupled locally to a bath of non-interacting conduction electrons through a spin-exchange interaction $J$. For an antiferromagnetic $J$, the impurity spin forms a singlet with the conduction band local spin density $S_0$ at sufficiently low temperatures. To investigate the impact of a local magnetic field on the Kondo effect, we consider the following model Hamiltonian:
\begin{equation}
  H = \sum_{k,\sigma} \varepsilon_k n_{k\sigma} + J \mathbf{S}_d \cdot \mathbf{S}_0 + B \mu_B \mathbf{S}_d^z~,
\end{equation}
where the first and second terms define the Kondo model and represent  the kinetic energy of the conduction bath and the impurity-bath spin interaction, respectively. The third term couples a local magnetic field to the impurity spin.

\subsection{Unitary Renormalisation Group (URG) Method}
To explore the low-energy phases of the externally field-dependent Kondo effect, we employ the Unitary Renormalization Group (URG) \cite{mukherjee2020holographic-A, mukherjee2020holographic-B}. The URG method is a scaling analysis for interacting fermions and quantum spins where high-energy degrees of freedom are systematically integrated out through the iterative application of many-particle unitary transformations. This approach unveils low-energy Hamiltonians comprised of emergent degrees of freedom and renormalized interaction couplings.

\begin{figure}[!ht]
\centering
\includegraphics[width=0.7\textwidth]{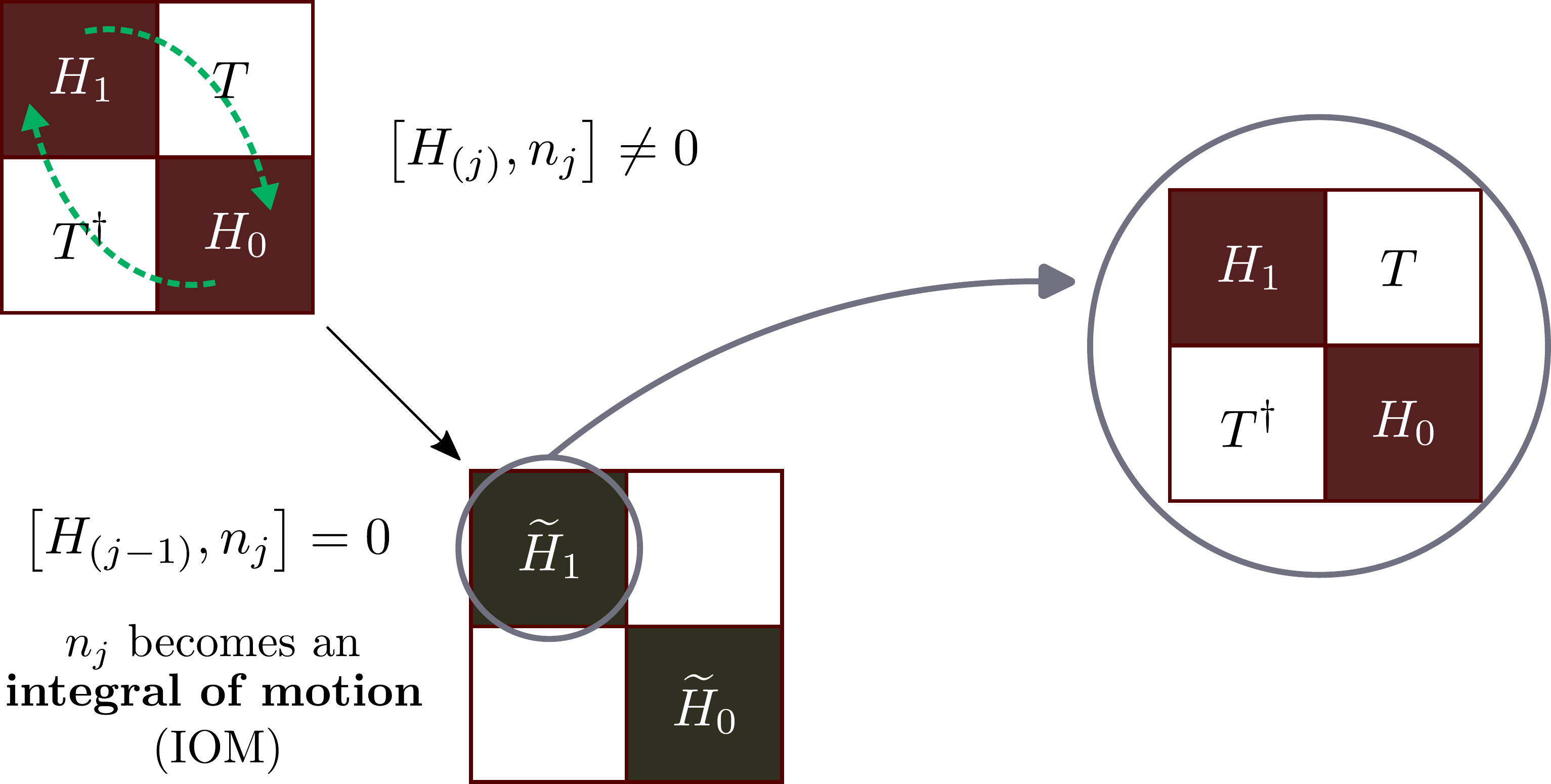}
\caption{Schematic diagram illustrating a single step of the URG process. One application of the unitary RG transformation decouples the fermionic states $n_j$ and modifies the couplings for the remaining states into new Hamiltonians $\tilde H_1$ and $\tilde H_0$.}
\label{Z2e}
\end{figure}

In the URG framework (see Fig.\ref{Z2e}), each step involves decoupling degrees of freedom by resolving quantum fluctuations, reducing the 
system's bandwidth from $D_N$ (the maximum bandwidth in the ultraviolet (UV) regime) to $D_{N-1}$ and continuing this process iteratively until reaching the low-energy bandwidth $D_0$ (the minimum bandwidth in the infrared (IR) regime). The decoupled degrees of freedom become integrals of motions (IOMs) under the RG evolution. At each step, the Hamiltonian transforms from $H_j$ to $H_{j-1}$, with $j$ representing the step index (starting from $N$ in the UV regime and ending at $0$ in the IR regime). As mentioned above, a single step of the URG process involves the change in the Hamiltonian by the application of a unitary transition $U_{j}$
\begin{equation}
H_{j-1} = U_j H_j U_j^\dagger~,
\end{equation}
where the unitary transformation $U_j$ can be expressed as
\begin{equation}
U_j = \frac{1}{\sqrt{2}} (1 + \eta_j - \eta_j^\dagger)~,
\end{equation}
and the operators $\eta_j$ and $\eta_j^\dagger$ are fermionic (i.e., $\eta_j$ and $\eta_j^\dagger$ obey fermionic anti-commutation relations), with
\begin{align*}
\eta_j^\dagger = \frac{1}{w - \text{Tr}(H_j n_j)} c_j^\dagger \text{Tr}(H_j c_j)~,
\end{align*}
and $w$ denotes the energyscale for the quantum fluctuations resolved by the unitary transformation.
\par
Figure \ref{Z2e} demonstrates a single step of the URG process. The Hamiltonian is first partitioned into different sectors with regards to the particular high energy degree of freedom to be decoupled: the particle sector ($H_1$), the hole sector ($H_0$), with off-diagonal terms ($T$ and $T^\dagger$) that couple the particle and hole sectors. After the application of the unitary transformation, the high-energy degree of freedom is decoupled, resulting in a Hamiltonian that is in diagonal form ($\tilde{H_1}$ and $\tilde{H_0}$) in the old basis. However, the Hamiltonian still contains off-diagonal terms in the new basis. We iterate the URG process until the desired low-energy space is obtained, i.e., either a fixed point is reached where the RG evolution stops, or all degrees of freedom till the Fermi energy are decoupled.

\section{Quantum phase transition driven by Kondo breakdown}\label{kondobreakdownsection}
                           
Performing the Unitary Renormalization Group (URG) transformation on the magnetic field-induced Kondo effect (details in Appendix \ref{A!w}), we derive the renormalization group (RG) equations for the antiferromagnetic Kondo coupling $J>0$ and the magnetic field $B$:
\begin{equation}
\frac{\Delta J}{\Delta D} = - J^2 \frac{\Tilde{w}}{\Tilde{w}^2 - \left(\frac{\mu_B B}{2}\right)^2} N(D)~,~\frac{\Delta B}{\Delta D} = -\frac{J^2}{4} \frac{B}{\Tilde{w}^2 - \left(\frac{\mu_B B}{2}\right)^2} N(D)~,
\end{equation}
where $\Tilde{w}=w-\frac{D}{2} +\frac{J}{4}$ and $\Tilde{w}^2=(\frac{\mu_B B}{2})^2$ defines the fixed point. Here, $N(D)$ represents the density of states, and $J^*$ and $B^*$ denote the fixed-point coupling for Kondo $J$ and magnetic field, respectively.

\begin{figure}[!ht]
\centering
\includegraphics[scale=0.4]{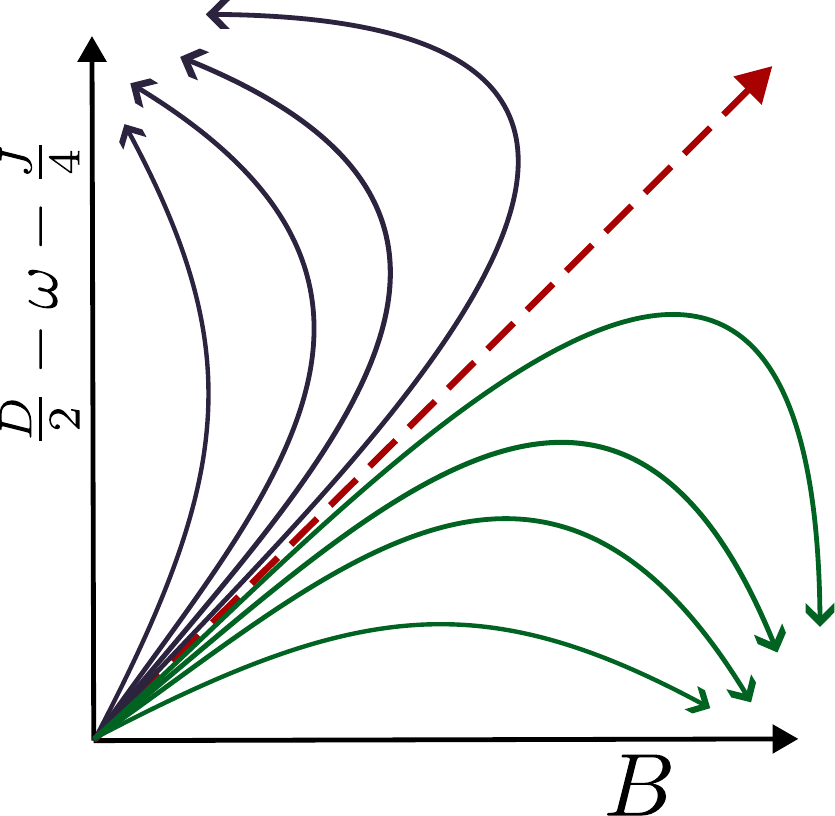}
\caption{Schematic RG phase diagram representing a measurement-induced quantum phase transition  between the Kondo screened (top left) and local moment (bottom right) fixed points of the Kondo problem in a local B-field.}
\label{o87}
\end{figure}

The competition between the coupling $J$ and the magnetic field $B$ in the RG equations given above can, in fact, be gleaned more compactly by noting that
\begin{equation}
\frac{\Delta B}{\Delta J} = \frac{B}{4 \Tilde{w}}~.\label{scale}
\end{equation}
We typically consider values of $w$ that are negative, making $\Tilde{w}$ either negative or positive depending on the competition between $\abs{w-\frac{D}{2}}$ and $\frac{J}{4}$. As shown schematically in Figure \ref{o87}, the relation between the two RG equations shown in eq.\eqref{scale} reveals the existence of a transition between a $J$ dominated phase and a $B$ dominated one.

\noindent\textbf{Kondo-screened strong-coupling phase:}\\
When $\abs{w-\frac{D}{2}} > \frac{J}{4}$  and $\Tilde{w}^2 > (\frac{\mu_B B}{2})^2 $, the coupling $J$ is relevant and the magnetic field $B$ is irrelevant. In this regime, the Kondo coupling $J$ is much stronger than the external magnetic field, preventing the magnetic field from destroying the associate local Fermi liquid metallic state~\cite{nozieres1974fermi}. The effective fixed-point Hamiltonian in this regime, considering negligible $B^*$ compared to $J^*$, is:

\begin{equation}
H_k = \sum_{|k|<\Lambda, \sigma} \varepsilon_k n_{k\sigma} + J^* \sum_{|k_1|, |k_2|<\Lambda} \mathbf{S}_d \cdot \mathbf{S}_{k_1 k_2}~.
\end{equation}

\begin{figure}[!ht]
\centering
\includegraphics[width=0.465\textwidth]{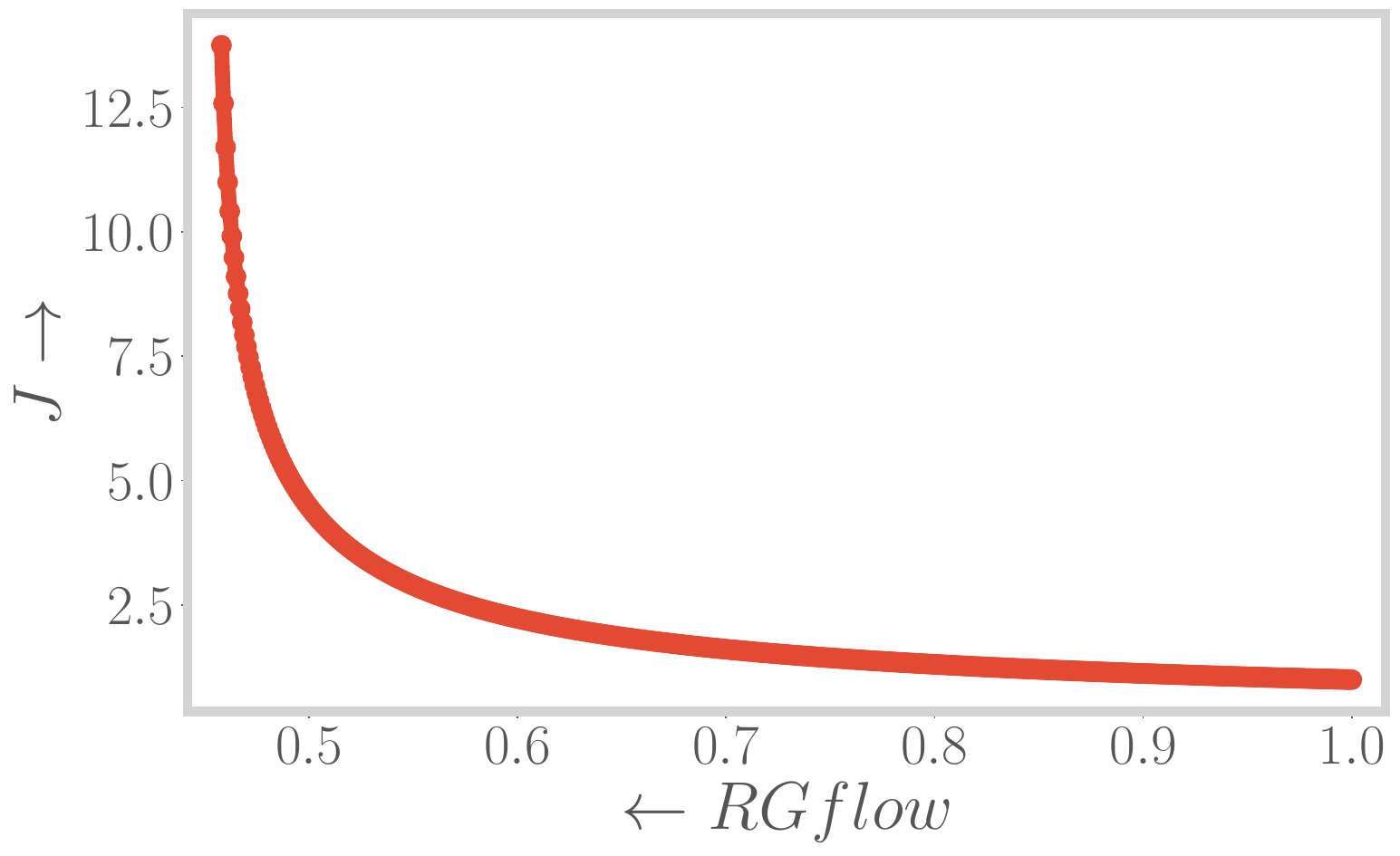}
\includegraphics[width=0.465\textwidth]{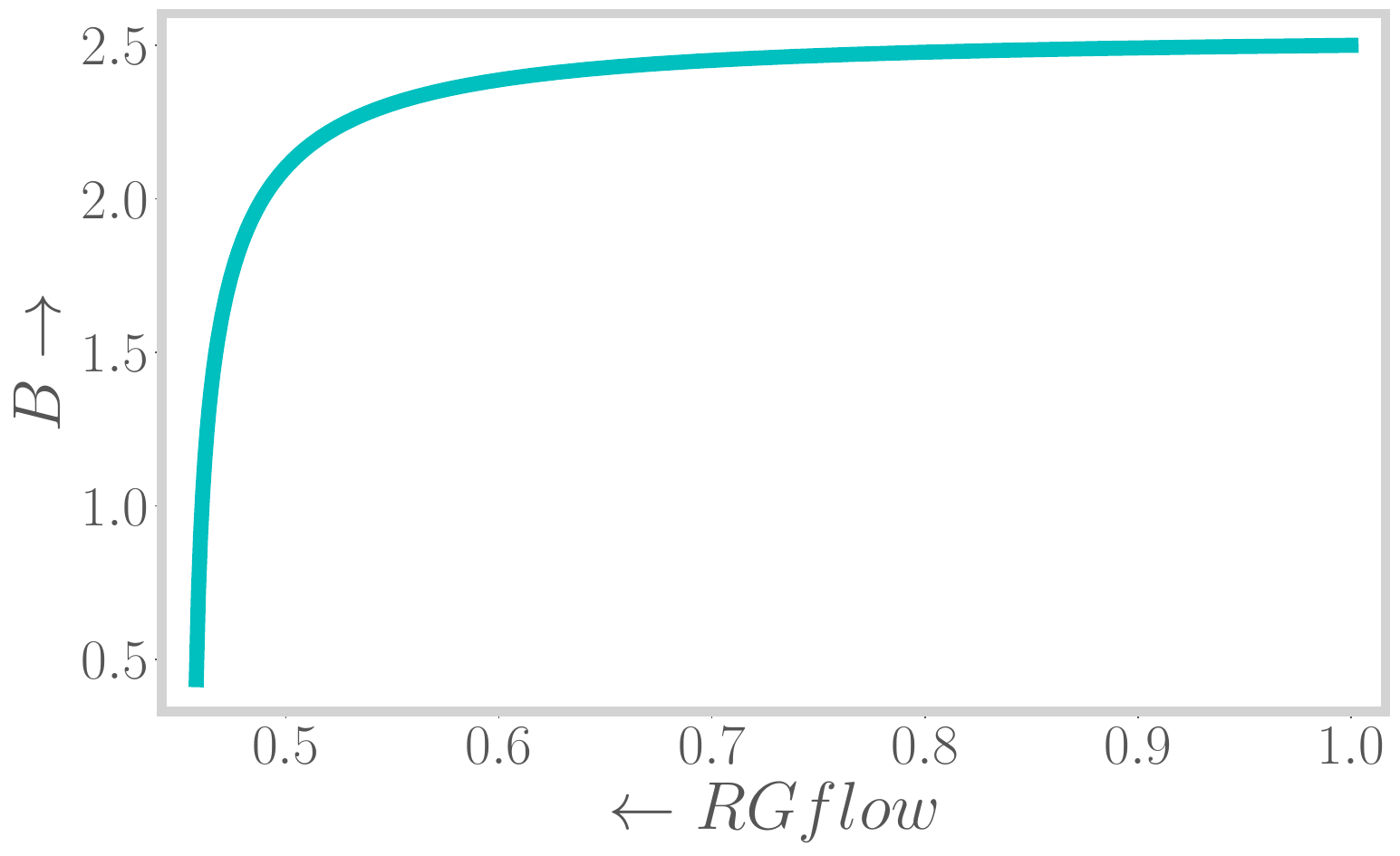}
\caption{RG flows showing relevant $J$ and irrelevant $B$ for a small external magnetic field in the Kondo screened regime.}
\end{figure}

\noindent \textbf{Local moment phase:}\\
When $\abs{w-\frac{D}{2}} > \frac{J}{4}$  and $\Tilde{w}^2 < (\frac{\mu_B B}{2})^2 $, the coupling $J$ becomes irrelevant, and the magnetic field $B$ is relevant. In this scenario, the Kondo singlet is destroyed, and the external magnetic field prevails, leading to the formation of a local moment (corresponding to a local insulating state, where the gapless excitations of the local Fermi liquid metal are gapped out). The effective Hamiltonian in this region, characterized by a decreasing $J$ due to a high $B$, is:
\begin{equation}
H_k = \sum_{|k|<\Lambda, \sigma} \varepsilon_k n_{k\sigma} + B^* \mu_B \mathbf{S}_d^z.
\end{equation}

\begin{figure}[!ht]
\centering
\includegraphics[width=0.465\textwidth]{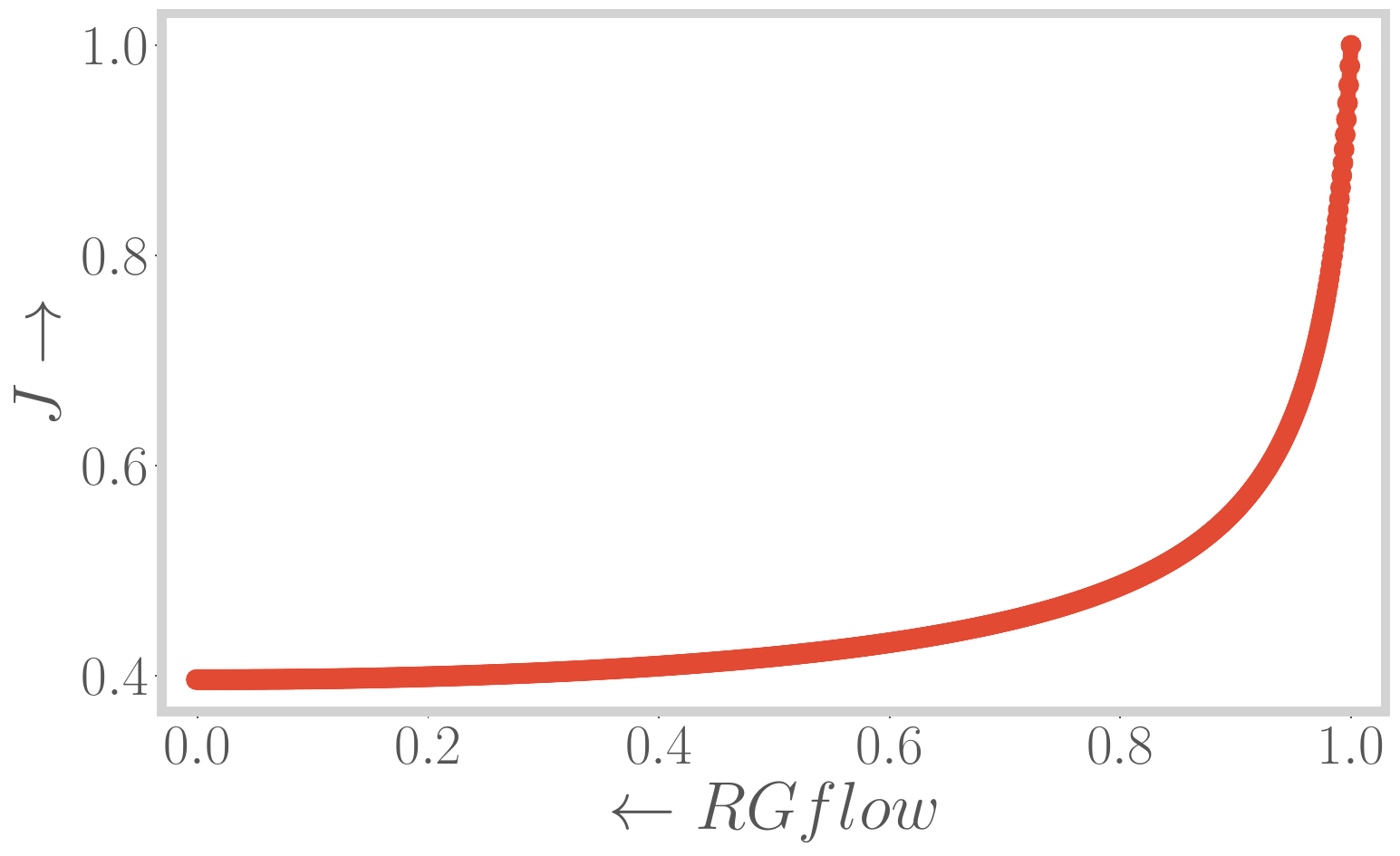}
\includegraphics[width=0.465\textwidth]{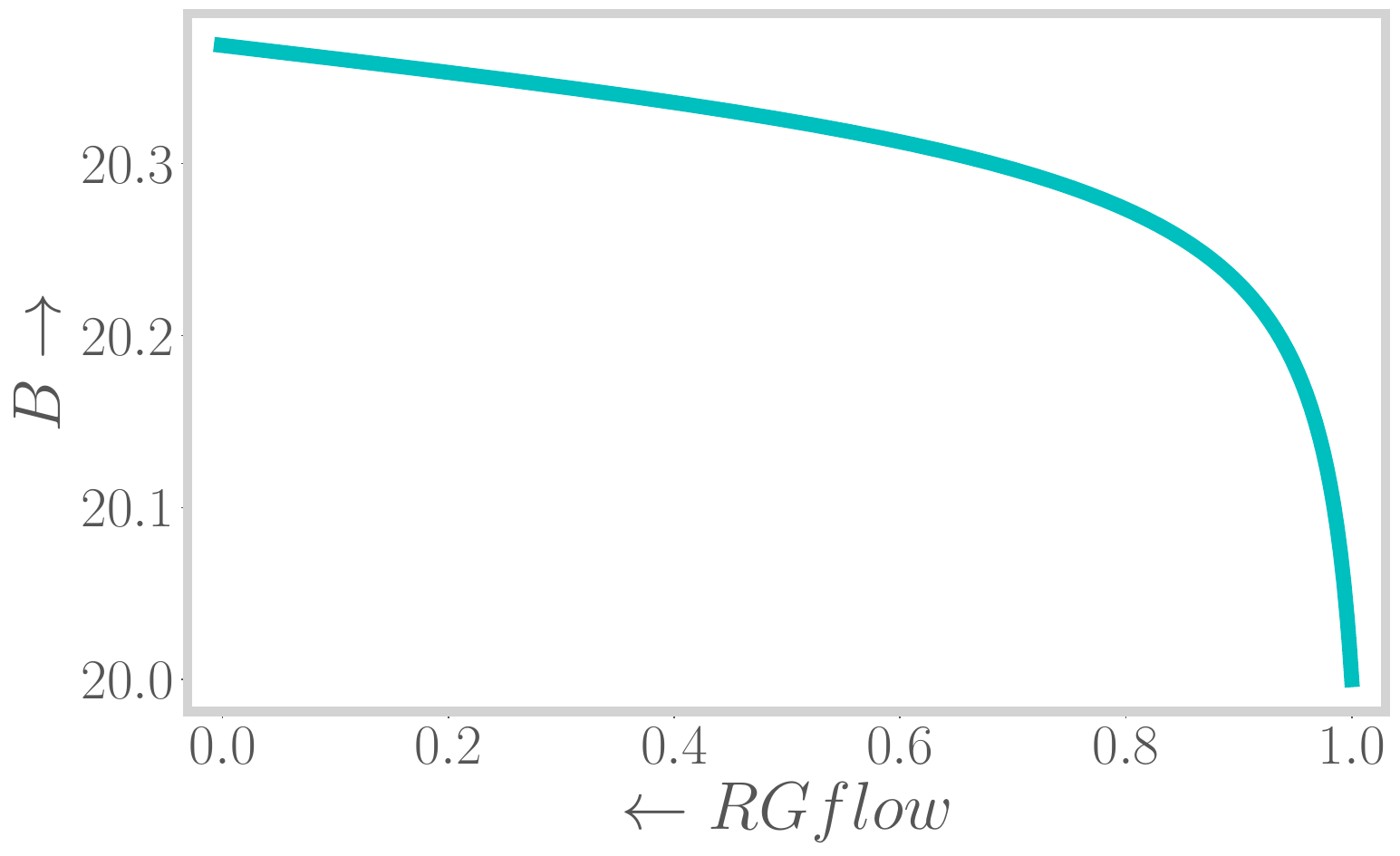}
\caption{RG flows showing irrelevant $J$ and relevant $B$ for the local moment regime.}
\end{figure}

\begin{figure}[!ht]
\centering
\includegraphics[width=0.6\textwidth]{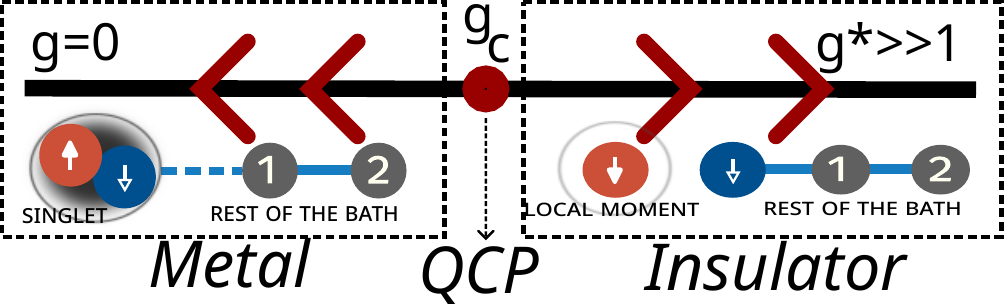}
\caption{Schematic diagram of the quantum-to-classical phase transition: the magnetic field induces a transition from the quantum singlet state to the classical localized state, where $g=\frac{\mu_B B}{J}$.}
\label{fig:d5}
\end{figure}

\noindent\textbf{Kondo breakdown transition:}
An external magnetic field induces a transition from singlet-backed local Fermi liquid metal to local moment insulator. A weaker magnetic field ($\Tilde{w}^2 > (\frac{\mu_B B}{2})^2 $) cannot overcome the strong Kondo coupling $J$, ensuring a metallic system. However, upon crossing the threshold magnetic field ($\Tilde{w}^2 = (\frac{\mu_B B}{2})^2 $), the Kondo coupling $J$ is rendered RG irrelevant. The singlet then destabilizes, and a local moment state forms via a quantum phase transition. The schematic diagram in Figure \ref{fig:d5} illustrates this transition due to tuning the magnetic field. This local quantum phase transition is also observed for the case of a conduction bath whose bandwidth is tuned towards the thermodynamic limit (Appendix \ref{S@3}). The magnetic field frustrates the maximally entangled singlet state, transforming it into a symmetry-broken polarized local moment state. In this sense, the phase transition represents a change from a state with quantum dynamics to one that is classical~\cite{jeong2014coarsening, naik2024tensile}: determining the ground state in the presence of a strong magnetic field becomes trivial. We also note in passing that the threshold local magnetic field given above for the transition vanishes for the case of a ferromagnetic Kondo interaction ($J<0$, that is marginally irrelevant under RG flow): this is a related to the fact that the dissipative coupling quenches quantum mechanical tunneling under RG flow in the corresponding spin-boson problem~\cite{guinea1985bosonization}.

\section{Nature of gapless excitations at the critical point}\label{criticalpointsection}\label{CritPeffH}
In order to understand the nature of the low energy theory that governs the transition of our system, we consider the effects of electron hybridisation between ground state of the zero-bandwidth Hamiltonian (describing the impurity and the local ``zeroth" site of the conduction bath that is coupled to it via the Kondo coupling, as discussed in Appendix \ref{WH3}) and a 1D conduction bath of electrons with a dispersion relation $\varepsilon_k = -2t \cos k$ and Fermi energy $\varepsilon_F=\mu$. We assume that the transition is attained at a certain finite value of the dimensionless coupling $g*=\frac{B^* \mu_B}{J^*}$, such that the distorted singlet ground state of the isolated zero-bandwidth Hamiltonian ($\ket{\Tilde{ss}}$) becomes degenerate with the polarised state $\ket{\downarrow_d \downarrow_0}$. 

The hybridisation term given by $V= -t \sum_{\sigma}(c_{0\sigma}^\dagger c_{1\sigma} + c_{1\sigma}^\dagger c_{0\sigma})$ (where the index $1$ refers to the next site of the conduction bath after the (zeroth) one that couples to the impurity through the Kondo coupling) is treated as a perturbation. Our goal is to determine whether the interaction of the degenerate ground states with the conduction electrons results in low-energy excitations that resemble those of a local Fermi liquid~\cite{nozieres1974fermi}, or lead to novel non-Fermi liquid behavior. At first order, the perturbation $V$ gives no contribution, $\bra{\psi} V \ket{\psi} =0$. Thus, we must consider the second-order perturbation, $VGV$, where $G$ is the Green's function given by:   $G=\frac{1}{E_i-E_n}$. Here, $E_i$ is initial energy and $E_n$ is the intermediate energy.

\par\noindent\textbf{Perturbation theory to second order}\\
In order to simplify the computation of the effective Hamiltonian, we denote the two degenerate states $\ket{\downarrow_d \downarrow_0}$ and the distorted singlet ($\ket{\Tilde{ss}}$) as $\ket{\Downarrow}$ and $\ket{\Uparrow}$ respectively. The effective Hamiltonian can then be written as
\begin{equation}
\mathcal{H} =  P_{n_{d,0}=1} \Big[\mathcal{J}^\perp \Big(S_{gs}^- S_1^+ + S_{gs}^+ S_1^- \Big) + \mathcal{J}^z S_{gs}^z S_1^z + \mathcal{H}_0^1 S_{gs}^z + \mathcal{H}_0^2 S_1^z \Big]~,
\label{effHam2}
\end{equation}
where the we have used the following operators:
\begin{equation}
S_{gs}^+ = \ket{\Uparrow} \bra{\Downarrow}~,~ 
S_{gs}^- = \ket{\Downarrow} \bra{\Uparrow}~,~
\ket{\Uparrow} \bra{\Uparrow} = P_{n_{d,0}=1} \left(\frac{1}{2} + S_{gs}^z \right)~,~
\ket{\Downarrow} \bra{\Downarrow} = P_{n_{d,0}=1} \left(\frac{1}{2} - S_{gs}^z \right)~.
\end{equation}
The spin operators $S_1^z$, $S_1^+$, $S_1^-$ follow the usual spin notation for the first site of the conduction bath. The analytical expressions of the coefficients $\mathcal{J}^\perp$, $\mathcal{J}^z$, $ \mathcal{H}_0^1$, and $\mathcal{H}_0^2$ are provided in Appendix \ref{W13}.

The effective Hamiltonian eq.\eqref{effHam2} corresponds to an anisotropic Heisenberg model for two effective spin-1/2 degrees of freedom. While in the absence of a magnetic field, the Kondo singlet formed between the impurity site and the 0th site of the conduction bath, eq.\eqref{effHam2} implies that at the transition, an effective singlet is formed which includes the first site of the conduction bath as well. Naturally, the bond between the impurity site and the 0th site must become weaker. The spin-flip fluctuation term in eq.\eqref{effHam2} is very different from the form of the local Fermi liquid (LFL) effective Hamiltonian obtained by Nozieres in the absence of the magnetic field~\cite{nozieres1974fermi}, and highlights the emergences of non-Fermi liquid (NFL) behavior precisely at the transition due to an orthogonality catastrophe arising from tunneling between the two degenerate ground states~\cite{si_kotliar_1993, varma2002singular, kotliarsi_1993, affleck2005non}.
   
\par\noindent\textbf{Perturbation theory to fourth Order}\\
Similar to the first order term, the terms are zero once again at third-order in perturbation theory. Therefore, we proceed to the computation of at fourth-order, and find: 
\begin{eqnarray}
&& \ket{\Downarrow} \bra{\Downarrow} \Biggl[ \alpha \ket{\uparrow_1} \bra{\uparrow_1} -\beta \ket{\downarrow_1 } \bra{\downarrow_1 } \Biggl] + \ket{\Uparrow} \bra{\Uparrow} \Biggl[ \gamma \Big( \ket{0_1} \bra{0_1} + \ket{\uparrow_1 \downarrow_1} \bra{\uparrow_1 \downarrow_1} \Big) + \eta \ket{\downarrow_1} \bra{\downarrow_1} \Biggl]\nonumber\\
&& + \zeta \Big(\ket{\Downarrow} \bra{\Uparrow} \ket{\uparrow_1} \bra{\downarrow_1} + \ket{\Uparrow} \bra{\Downarrow} \ket{\downarrow_1} \bra{\uparrow_1}\Big)
\end{eqnarray}
The analytical expressions for the coefficients $\alpha$, $\beta$, $\gamma$, $\eta$ and $\zeta$ are provided in the Appendix \ref{D&5}. 
The set of terms in the first big bracket indicates Zeeman splitting of spins, as the coefficients $\alpha$ and $\beta$ are different. The set of terms in the second big bracket connects adiabatically to that of a LFL~\cite{nozieres1974fermi} repulsion at the first site as the external magnetic field goes to zero (i.e., $\eta$ vanishes)~\cite{mukherjee2022unveiling, nozieres1980kondo}. Finally, the last the terms indicate, similar to the spin-flip terms obtained at second order, tunneling between the two degenerate ground states that likely result in non-Fermi liquid (NFL) behaviour. We provide further evidence for the breakdown of the LFL in the next section.
    
\section{Spectral properties and eigenstate thermalisation of impurity spin}\label{spectraltherm}
\subsection{Spectral Function and Quasiparticle Residue}
At zero temperature, the impurity spectral function is given by $A(\omega) =-\frac{1}{\pi} Im[G_{dd}^\sigma(\omega)]$ where $G_{dd}^\sigma(\omega)$) is the impurity retarded Green’s function, $G_{dd}^\sigma(\omega) = -i \theta(t) \langle \{\mathcal{O}_\sigma(t), \mathcal{O}_\sigma^\dagger \} \rangle$. Here, the operator $\mathcal{O}_\sigma = S_d^{-\sigma} c_{0, -\sigma}$ corresponds to the low-energy excitation of the Kondo singlet. In terms of the eigenstates of the impurity model, spectral function takes the form:
\begin{equation}
A(\omega) = \frac{1}{d_0} \sum_{n,0} \Big[ |\langle 0|\mathcal{O}_\sigma|n\rangle |^2 \delta (\omega + E_0 - E_n) +  |\langle n|\mathcal{O}_\sigma|0\rangle |^2 \delta (\omega - E_0 + E_n)\Big]~,
\end{equation}
where $d_0$ indicates the degeneracy, the index $0$ the ground state, $E_0$ the corresponding ground state energy, and $n$ and $E_{n}$ represents all the excited states and their corresponding energies. This spectral function is computed along the URG flow from UV to IR.

\begin{figure}[!ht]
\centering
\includegraphics[scale=0.6]{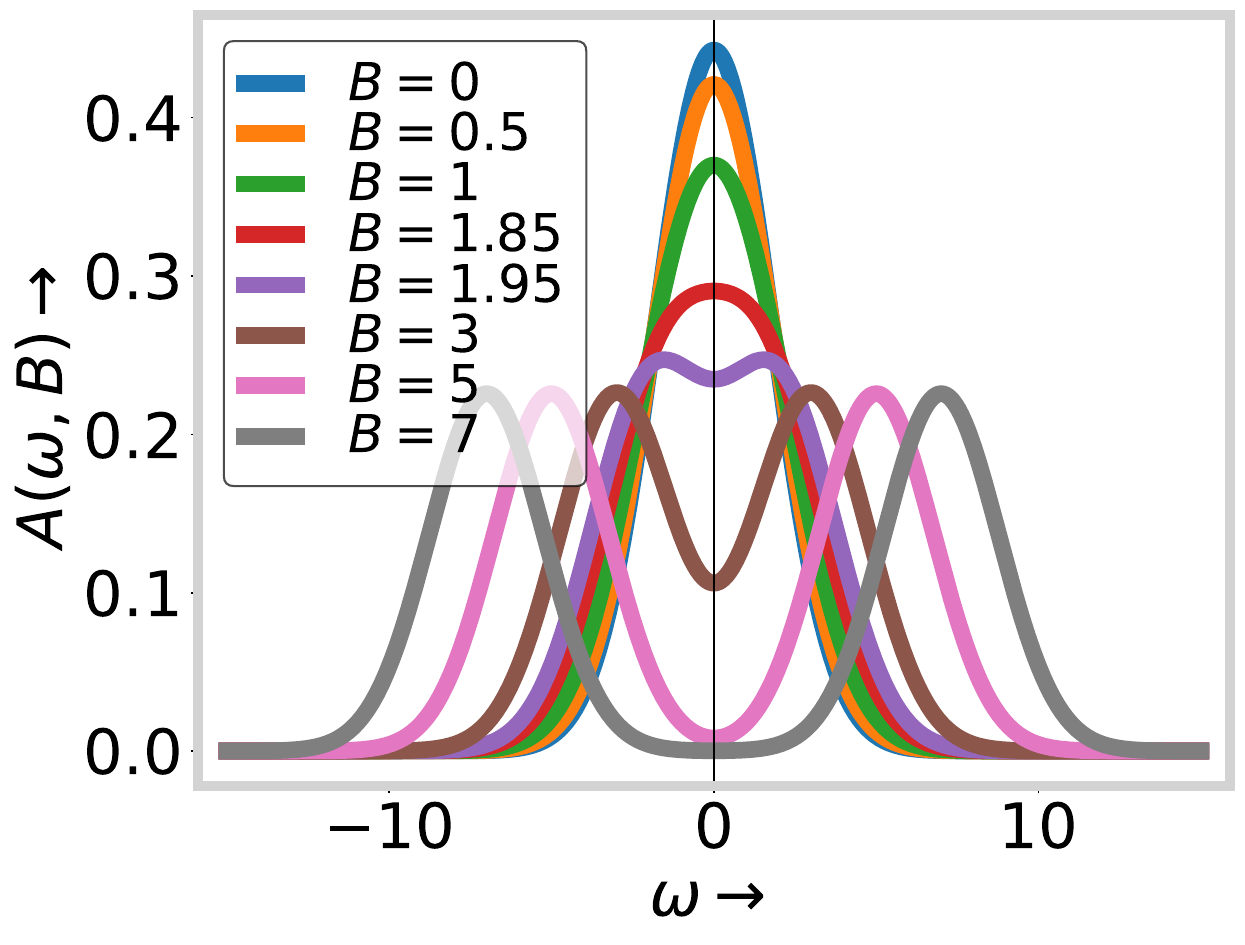}
\caption{Splitting of the impurity spectral function with increasing local B-field.}
\label{kjji}
\end{figure}

Figure \ref{kjji} displays the destruction of the Kondo peak as the magnetic field is increased. The peak splits~\cite{costi2000}, with the transition into the unscreened phase at the critical magnetic field $B^{*}=1.9$ for bare values of $J=0.2D$ and the bandwidth $D=1$, as dictated by the RG equations for $J$ and $B$ discussed above. The Kondo temperature ($T_k$) can be calculated from the half width-half maximum (HWHM) at $T=0$. We computed the magnetic field-dependent $T_k$ at $T = 0$ from the HWHM of the central Kondo peak. Following Ref.\cite{domanski2016constructive}, we also derived analytically $T_k$ from the local magnetic field-induced single impurity Anderson model (B-SIAM) in Appendix \ref{D4E}: 
\begin{equation}
	T_k = \frac{U_d}{2} \sqrt{\frac{16 U_d V^2 \rho}{U_d^2 - B^2}} \hspace{1mm} e^{-\frac{U_d^2 - B^2}{16 U_d V^2 \rho}}~.
\end{equation}

\begin{figure}[!ht]
\centering
\includegraphics[scale=0.31]{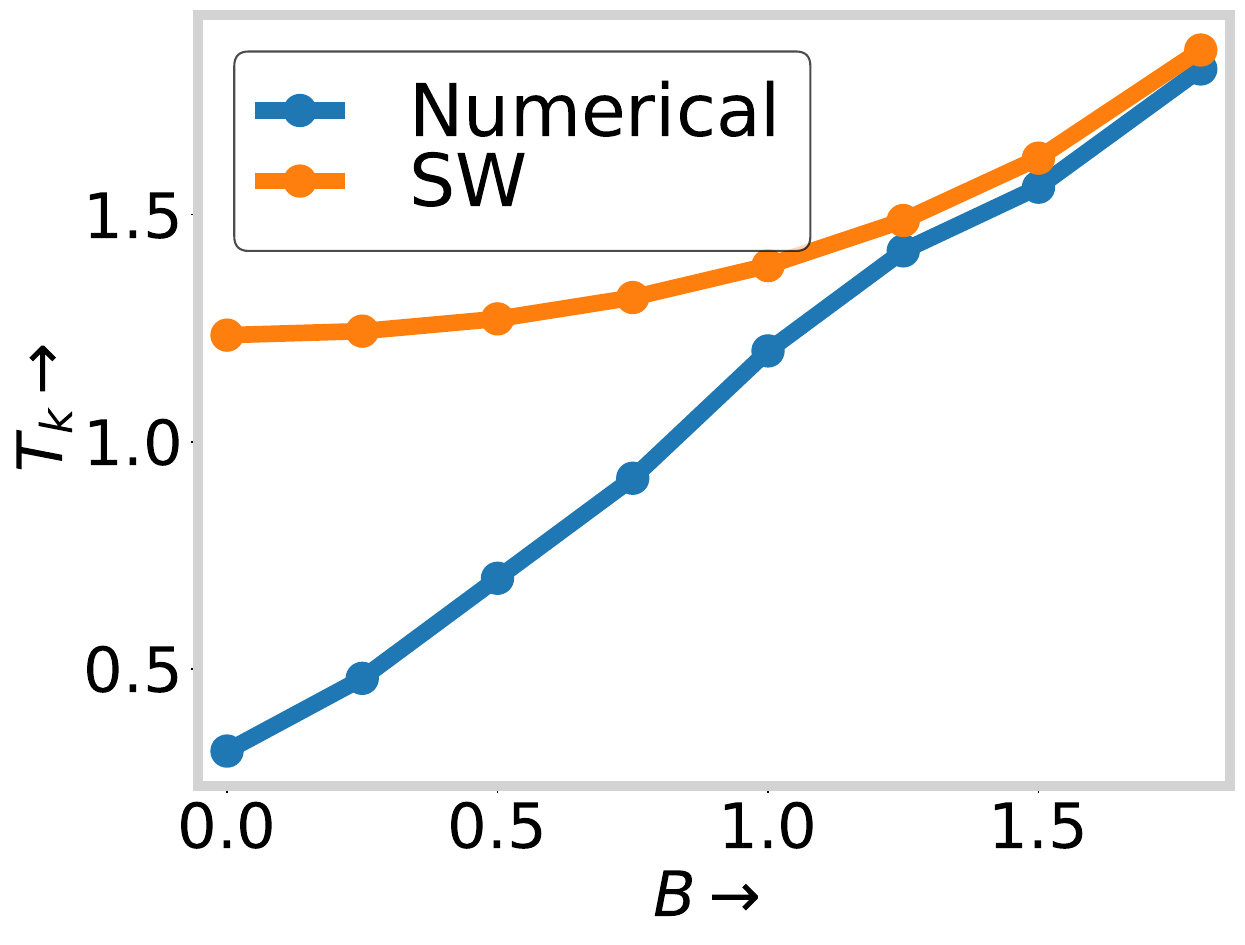}
\includegraphics[scale=0.31]{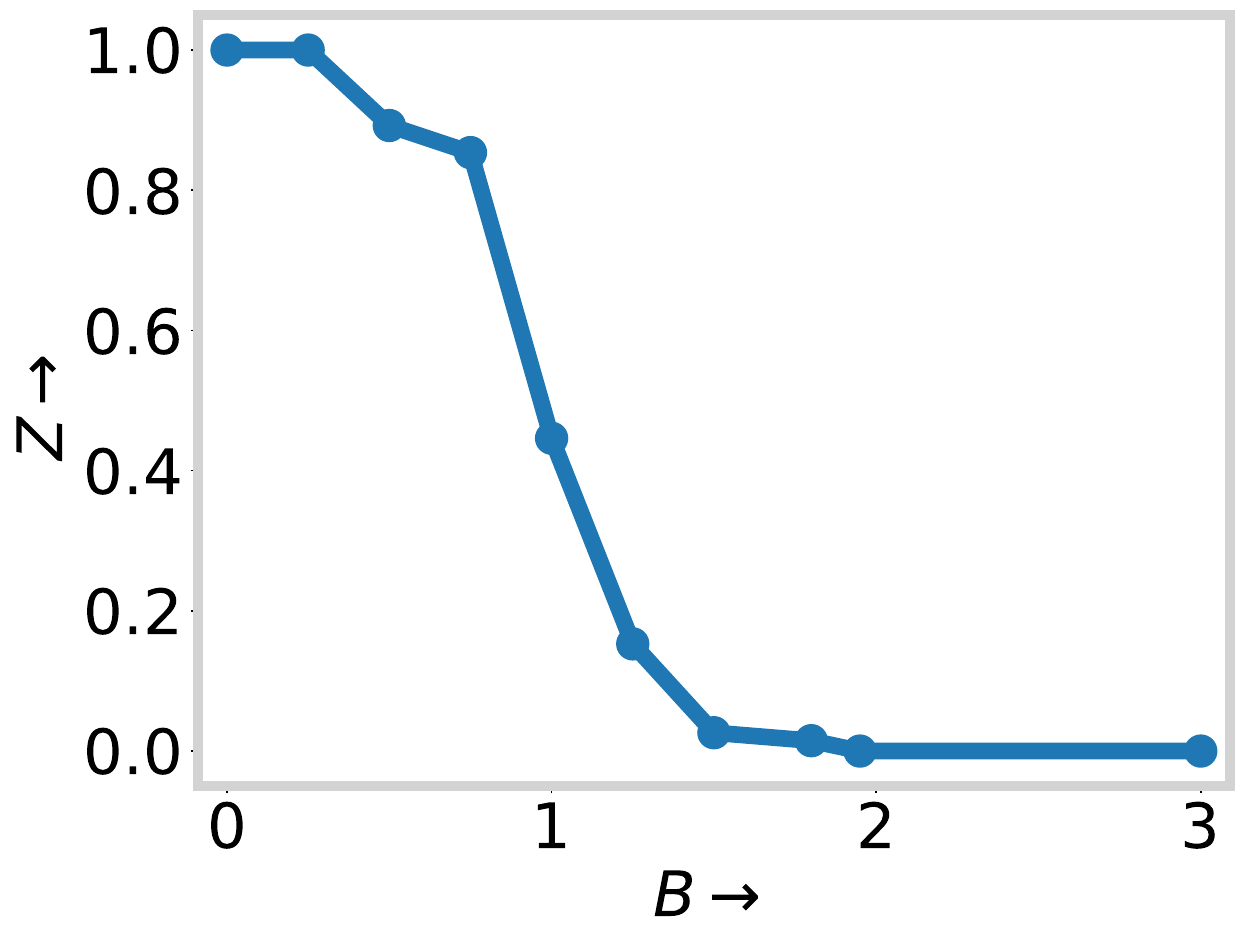}
\caption{Magnetic field dependence of (left) Kondo Temperature $T_{k}$ and (right) quasiparticle residue ($Z$).}
\label{nkj}
\end{figure}

In Fig.\ref{nkj}(a), we plot both the analytical expression for the Kondo temperature $T_{k}$ given above as well as that obtained numerically from the URG flow. For the analytic expression, we used the values $U_d = 3$, $V= 0.6$ and $\rho = 1$. It is evident that the analytic expression matches the numerically obtained values better at high magnetic field. We expect that the poor agreement at low field arises from the fact that the analytic expression is obtained from a Schrieffer-Wolff transformation~\cite{schrieffer1966} (see Appendix \ref{D4E} for a detailed derivation): 
this corresponds to a single-shot RG process that misses the dynamical spectral weight transfer captured by our unitary etarative RG as the magnetic field is tuned. In Fig.\ref{nkj}(b), we plot the quasiparticle residue ($Z$) of the LFL metal (defined in terms of the area under the central Kondo peak in the impurity spectral function). The vanishing of $Z$ as $B$ is tuned towards the transition is an important indicator of the breakdown of Kondo screening, i.e., the loss of the quasiparticle excitations of the LFL and their replacement by the NFL excitations discussed earlier. We also present a detailed fit of the up-spin impurity spectral height in Appendix \ref{L8S}, and obtain an analytical expression for the $B$-field dependence of the impurity scattering phase shift $\delta_{\sigma}(B)$. 
    
\subsection{Breakdown of eigenstate thermalisation in the unscreened phase}    
\begin{figure}[!ht]
\centering
\includegraphics[scale=0.26]{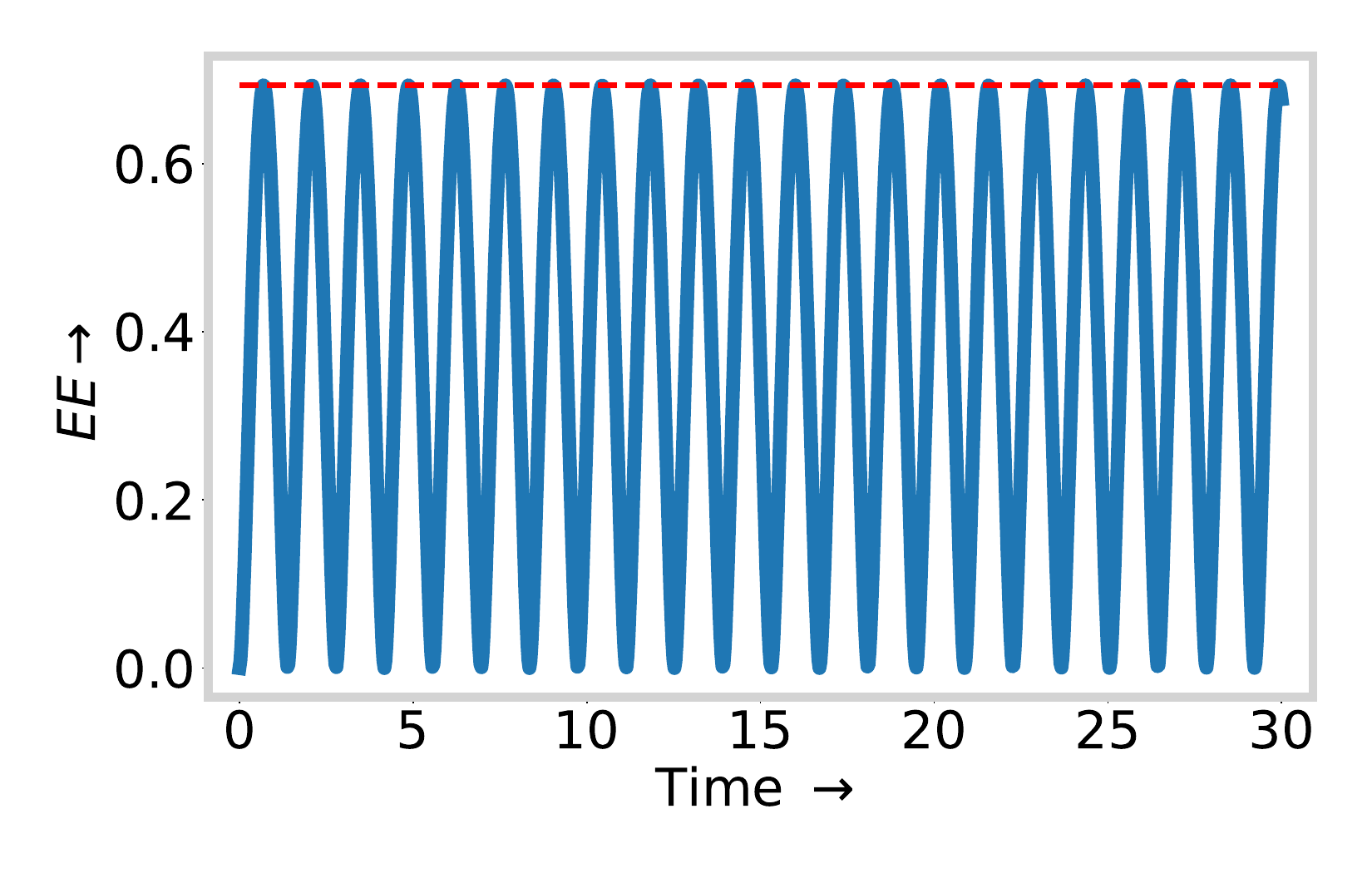}
\includegraphics[scale=0.26]{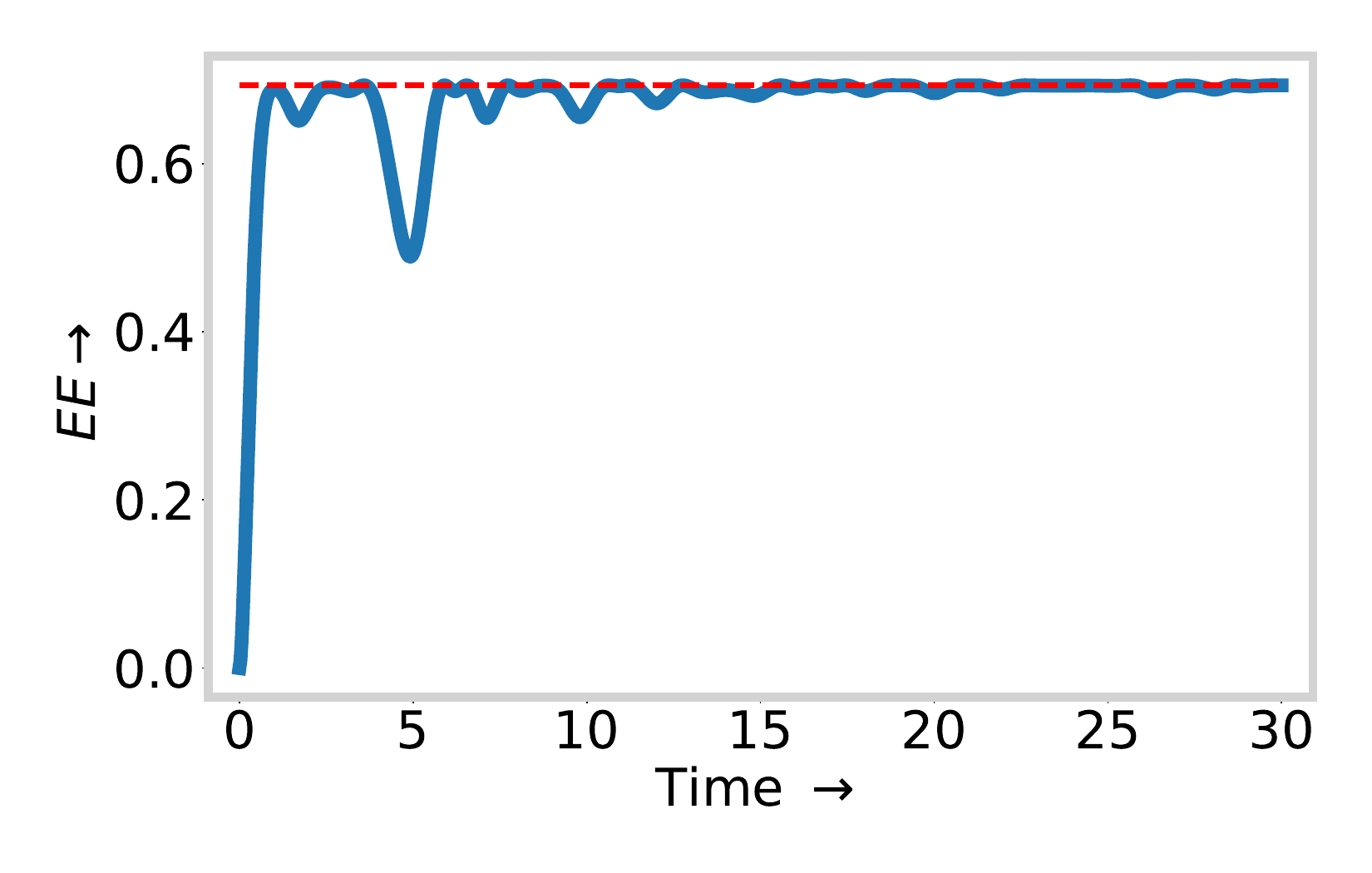}
\caption{Entanglement Entropy(EE) for (left) number of bath sites (left) $N_{\text{bath}}$ = 1 and (right)) $N_{\text{bath}}$ = 7(right).}
\label{HBG}
\end{figure}

\begin{figure}[!ht]
\centering
\includegraphics[scale=0.26]{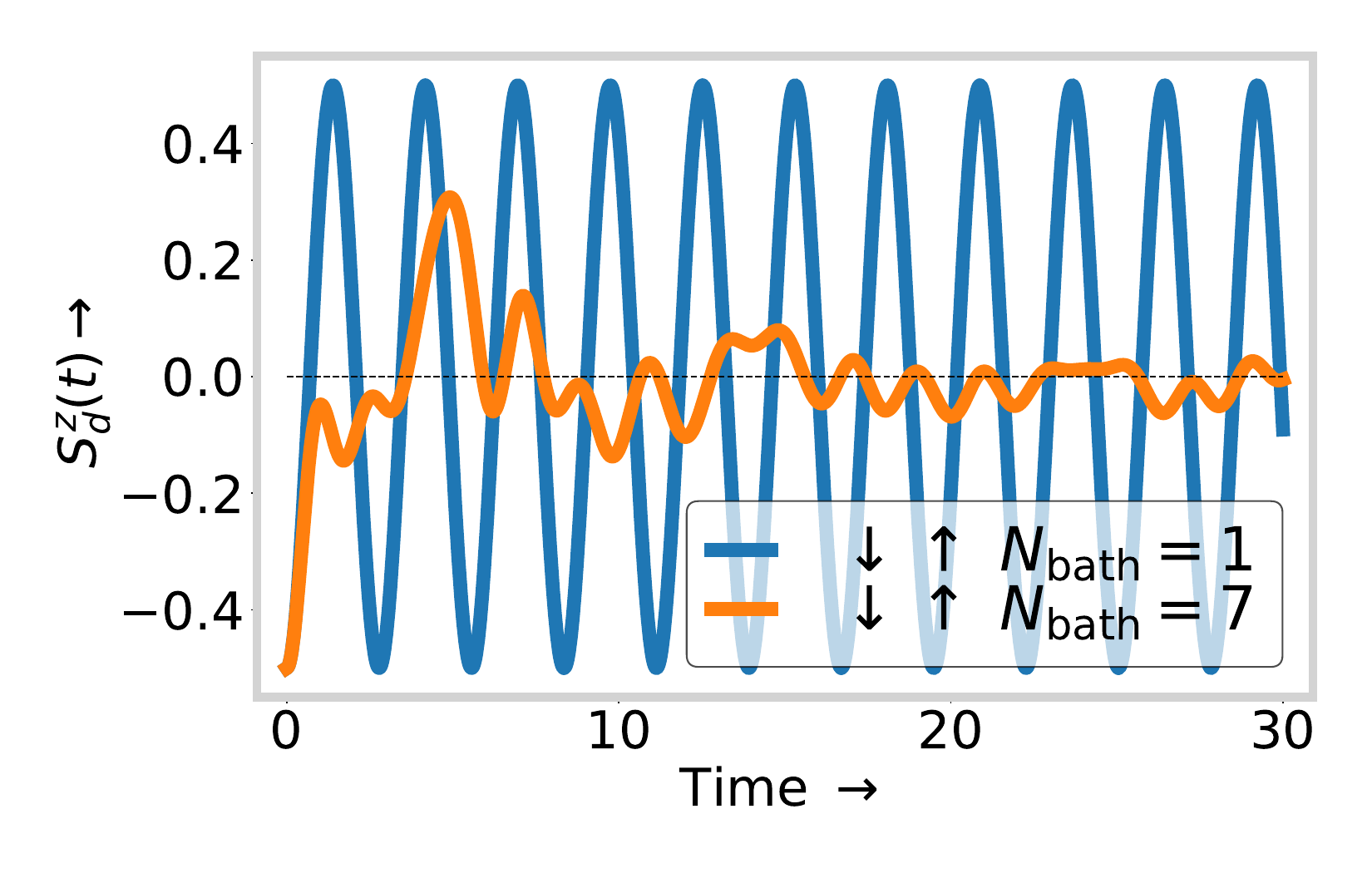}
\includegraphics[scale=0.26]{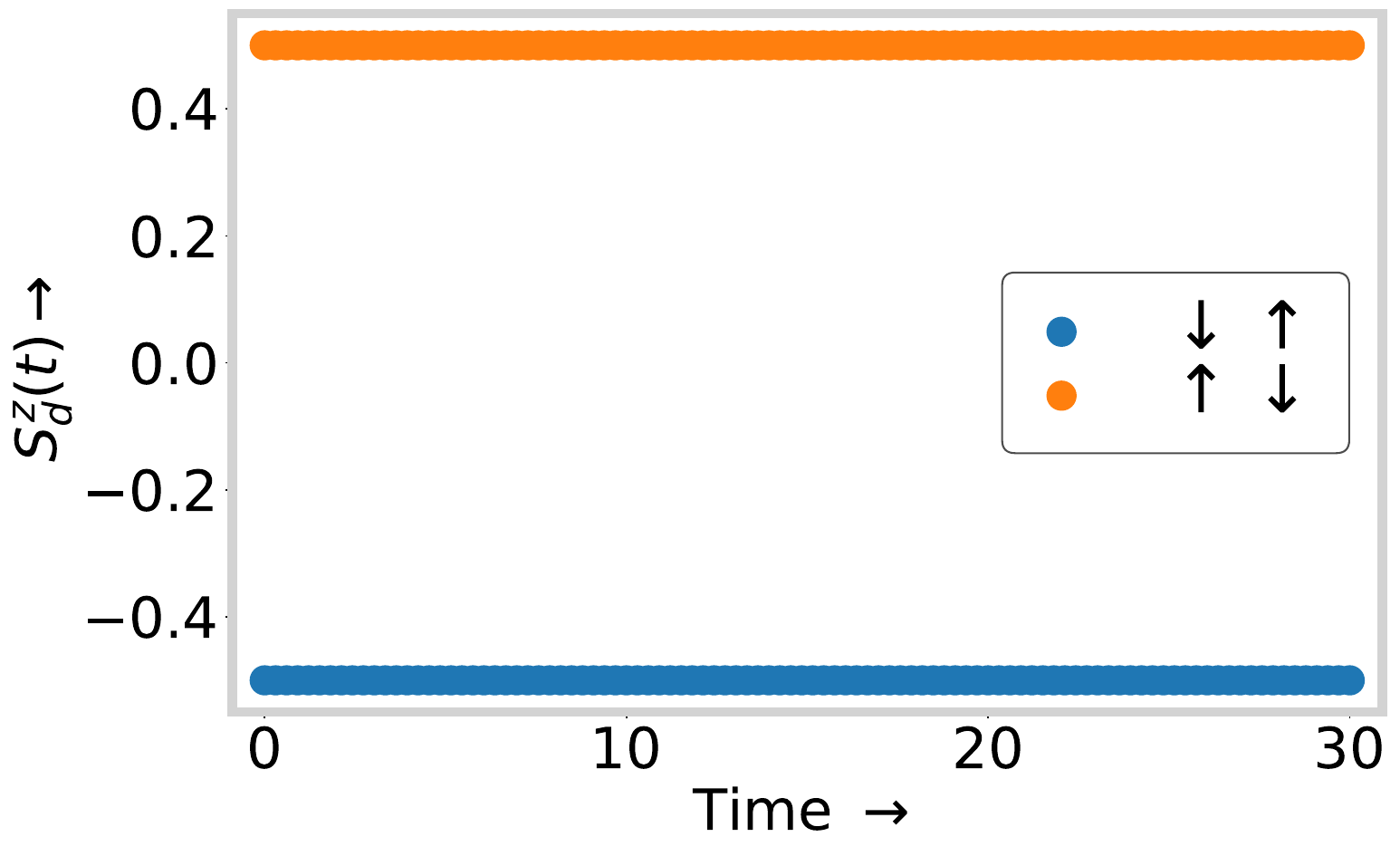}
\caption{Time evolution of $S_d^z$ for (left) metallic phase and (right) local moment phase.}
\label{DCF}
\end{figure}

Thermalization can be observed in the quantum dynamics of our system by starting from a non-eigenstate configuration (i.e., where the impurity spin is down and the bath spins alternate between up and down (up, down, up, down, ...) as the number of bath sites increases in the LFL metallic phase. This is a signature of the decoherence-induced growth of entanglement between the impurity spin and it's fermionic environment in this phase. Thus, the entanglement entropy of the impurity can be used to quantify the amount of decoherence. As shown in Fig. \ref{HBG} (b), the entanglement entropy (EE) of this phase increases and stabilizes at \( \log 2 \) as the number of baths sites ($N_{bath}$) is increased, indicating that the system reaches the maximally entangled singlet state upon tuning $N_{bath}$. This behavior underscores the role of bath size in driving the system towards the thermalization of the chosen initial state. Additionally, the oscillatory behavior of the observable \( S_d^z \) shown in Fig. \ref{DCF}(a) gradually diminishes and eventually saturates at 0 as the system size is increased. In contrast, in the local moment phase, no thermalization obtains: the chosen initial state remains frozen at \( -0.5 \) regardless of the bath size. If all the spins (both impurity and bath) are flipped, the system freezes at \( 0.5 \) (Fig. \ref{DCF}(b)). 

\section{Evolution of local observables during UV-IR crossover}\label{mergsection}
In this section, we employ the momentum space entanglement renormalization group (MERG) method~\cite{mukherjee2022unveiling} to observe the RG evolution of various local observables and entanglement measures related to the impurity in both Kondo-screened  and local moment phases of the impurity spin. In MERG, we start from the low-energy (infrared, IR) state corresponding to the fixed point theory, and progress towards the high-energy (ultraviolet, UV) state through an iterative application of the inverse unitary transformations of the URG. For this, we start from the IR fixed-point Hamiltonian and diagonalize it to obtain the ground state \(\ket{\psi_0}\). Next, we apply the inverse of the unitary transformation from the URG to couple a hitherto decoupled degree of freeedom lying outside the IR (corresponding to an integral of motion (IOM) generated from the URG flow). This is represented as $U^\dagger \ket{\psi_0} = \ket{\psi_1}$, where
\begin{equation}
U^\dagger = 1 + \frac{1}{\omega - H_D} \sum_k \left( S_d^z \beta c_{k\beta}^\dagger c_{q\beta} + c_{d\beta}^\dagger c_{d\bar{\beta}} c_{k\bar{\beta}}^\dagger c_{q\beta} \right) +\frac{1}{\omega - H_D} \sum_k \left( S_d^z \beta c_{q\beta}^\dagger c_{k\beta} + c_{d\bar{\beta}}^\dagger c_{d\beta} c_{q\beta}^\dagger c_{k\bar{\beta}} \right)~,    
\end{equation}
and $H_D$ correspond to the diagonal terms of the Hamiltonian (described in Appendix \ref{A!w}).

We proceed to the UV space \(\ket{\psi_N}\) by iteratively coupling IOMs at every step. Various local measurables (with corresponding operators $\hat{O}$ ) are computed at each step: \(\bra{\psi_j} \hat{O} \ket{\psi_j}\), as are several entanglement measures obtained by appropriately tracing out the required degrees of freedom.

\begin{figure}[!ht]
    \centering
    \includegraphics[scale=0.75]{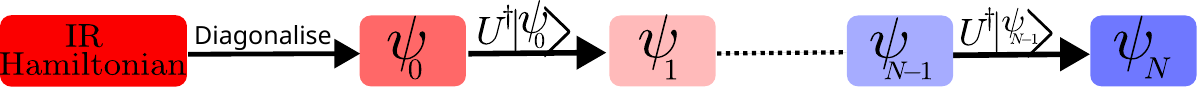}
    \caption{Schematic diagram of the MERG algorithm from IR to UV.}
\end{figure}

\begin{figure}[!ht]
    \centering
    \includegraphics[scale=0.27]{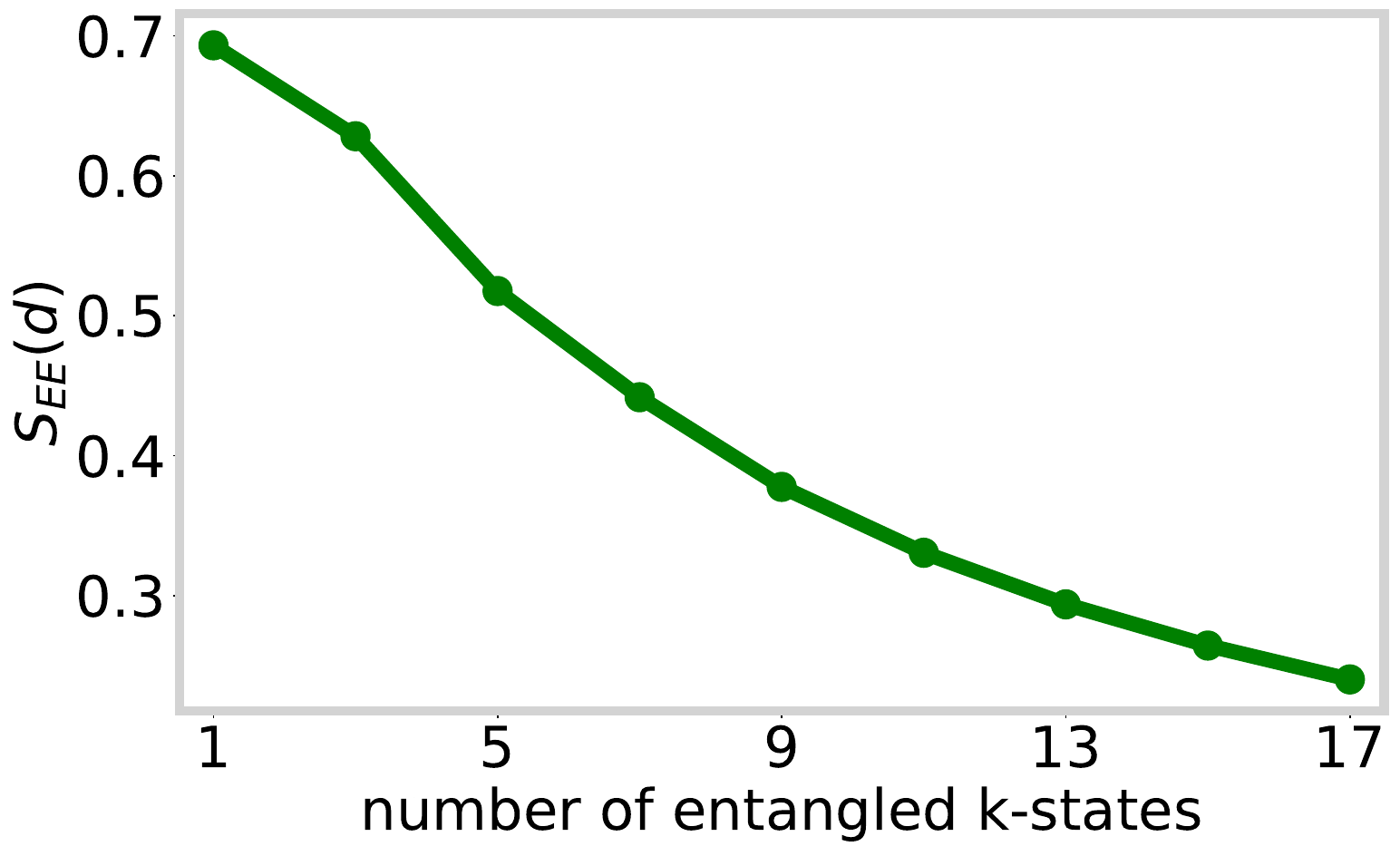}
    \includegraphics[scale=0.27]{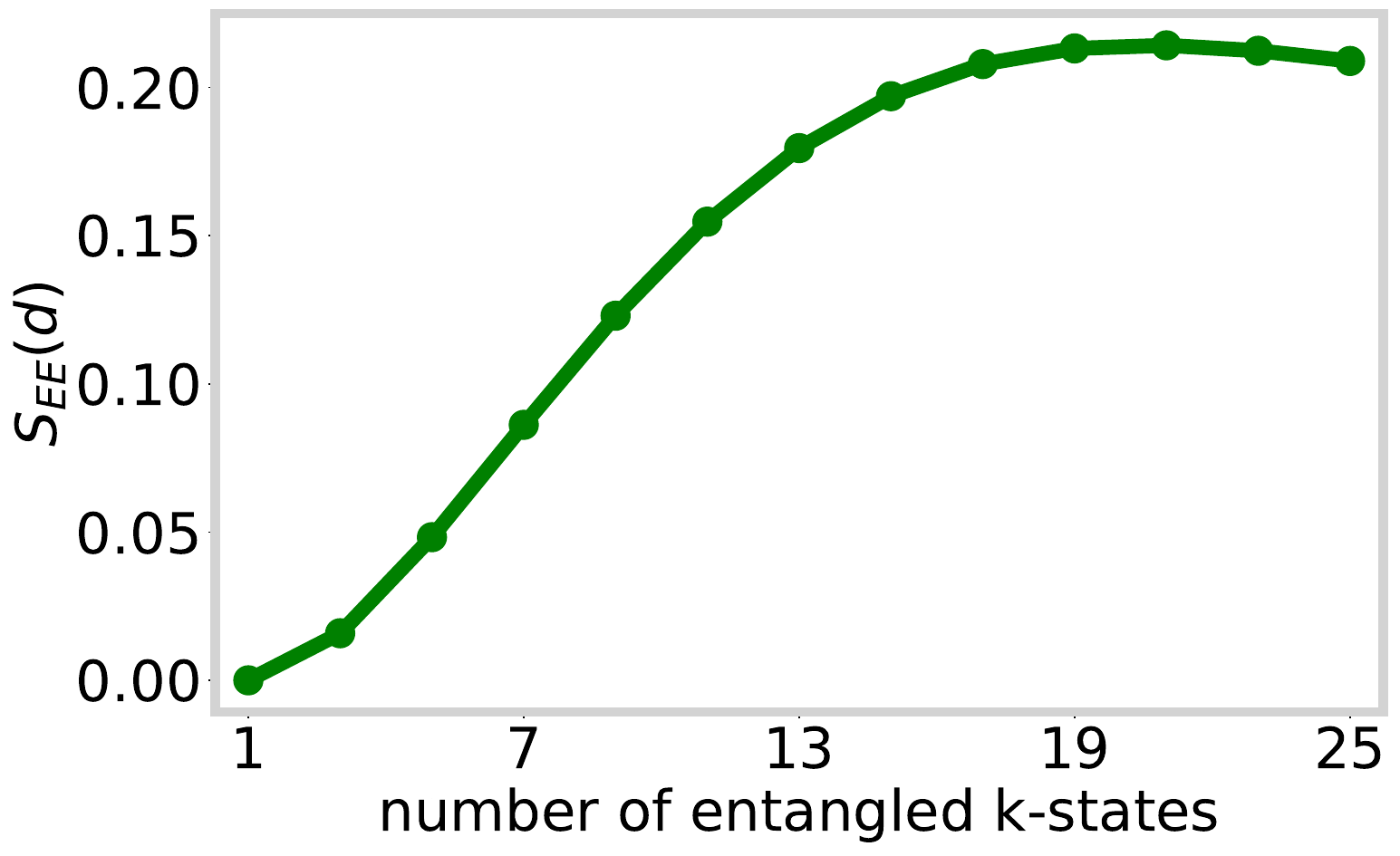}
    \caption{MERG evolution of Impurity Entanglement Entropy $S_{EE}(d)$ in (left) singlet regime and (right) local moment regime.}
    \label{A!2}
\end{figure}

We show the entanglement entropy of the impurity (\(S_{EE}(d)\)) in the metallic phase in Fig.~\ref{A!2}(a): the singlet region at the IR is clearly \(\ln 2\), while \(S_{EE}(d)\) decreases steadily with each step taken towards the UV. On the other hand, as shown in Fig.~\ref{A!2}(b) it is zero for the local moment region in the IR (as expected), and starts growing with each step taken towards the UV.

\begin{figure}[!ht]
    \centering
    \includegraphics[scale=0.26]{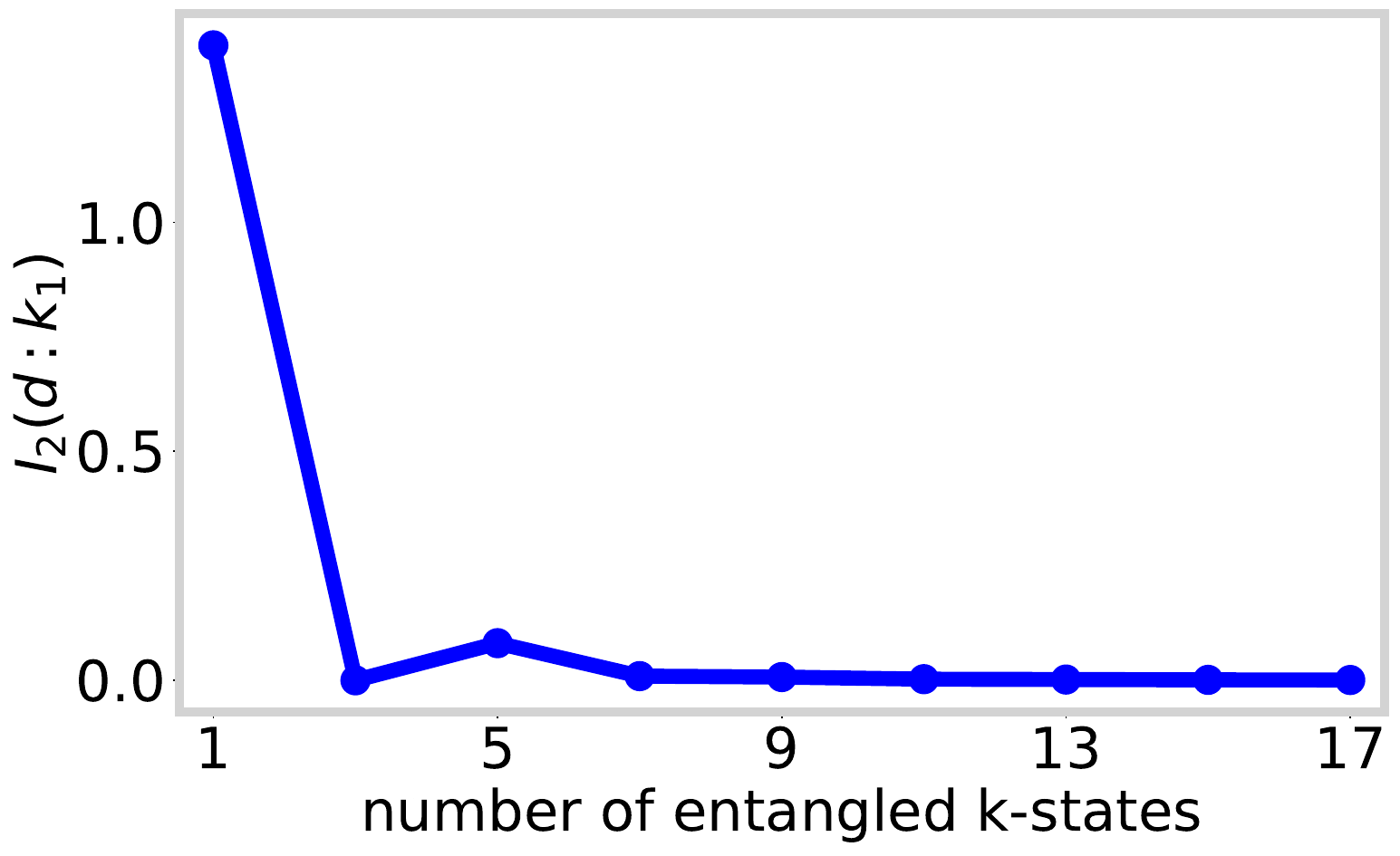}
    \includegraphics[scale=0.26]{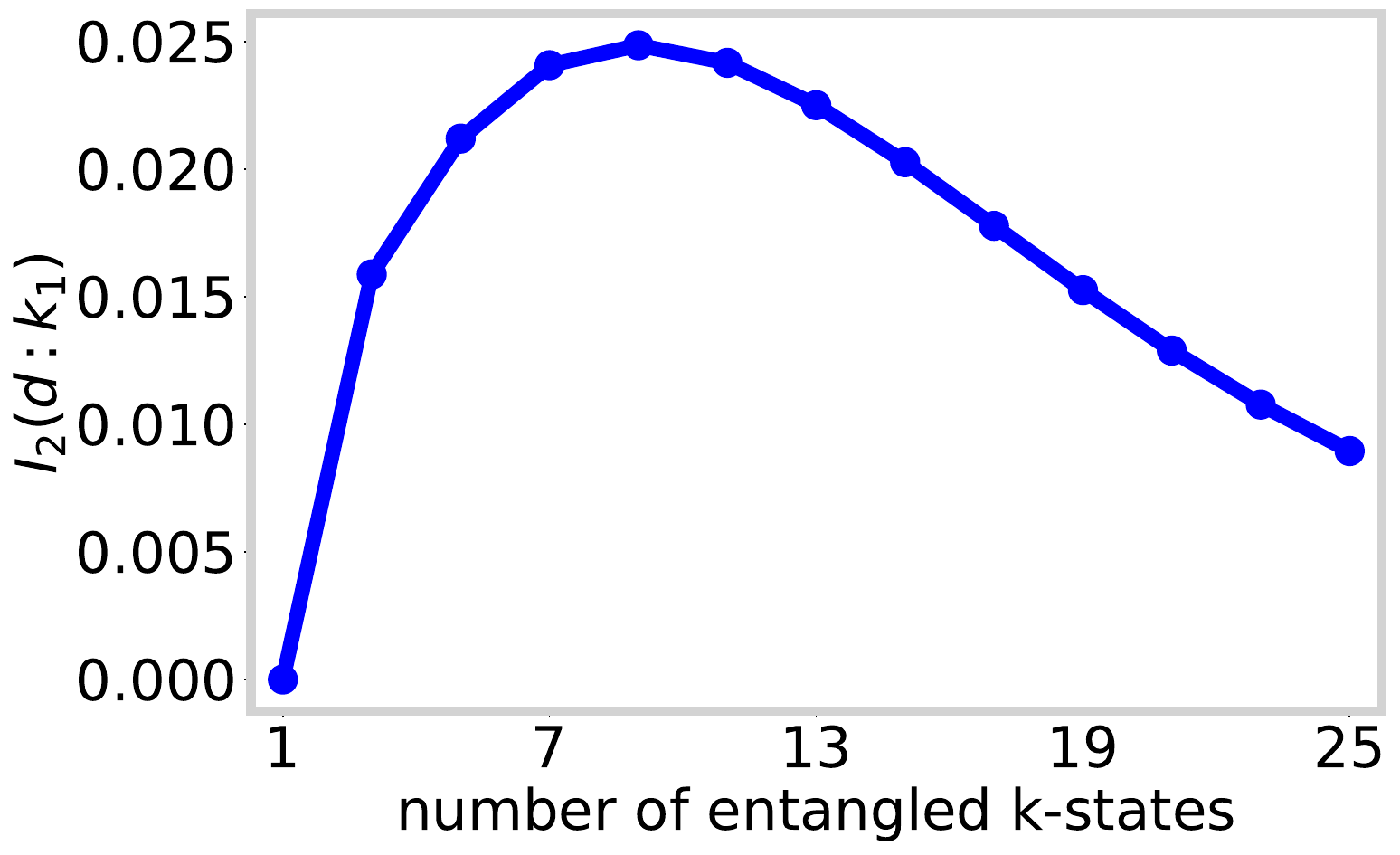}
    \caption{MERG evolution of Mutual Information between impurity and bath zeroth site $I_{2}(d:k_{1})$ in (left) singlet regime and (right) local moment regime.}
    \label{mutinfo}
\end{figure}

The bipartite mutual information between two subsystems $A$ and $B$ is defined as $I_2(A:B) = S_{EE}(A) + S_{EE}(B) - S_{EE}(A \cup B)$. As shown in Fig.\ref{mutinfo}, the mutual information $I_{2}(d:k_{1})$ between the impurity (\(d\)) and the first \(k\) site (where \(k\) is the Fourier transform of the 0-th site) in the metallic side increases under RG flow from UV to IR (similar to Ref.\cite{mukherjee2022unveiling}). On the insulating local moment side, $I_{2}(d:k_{1})$ is small throughout the RG evolution and shows a non-monotonic behaviour in flowing from UV to IR.

\begin{figure}[!ht]
    \centering
    \includegraphics[scale=0.26]{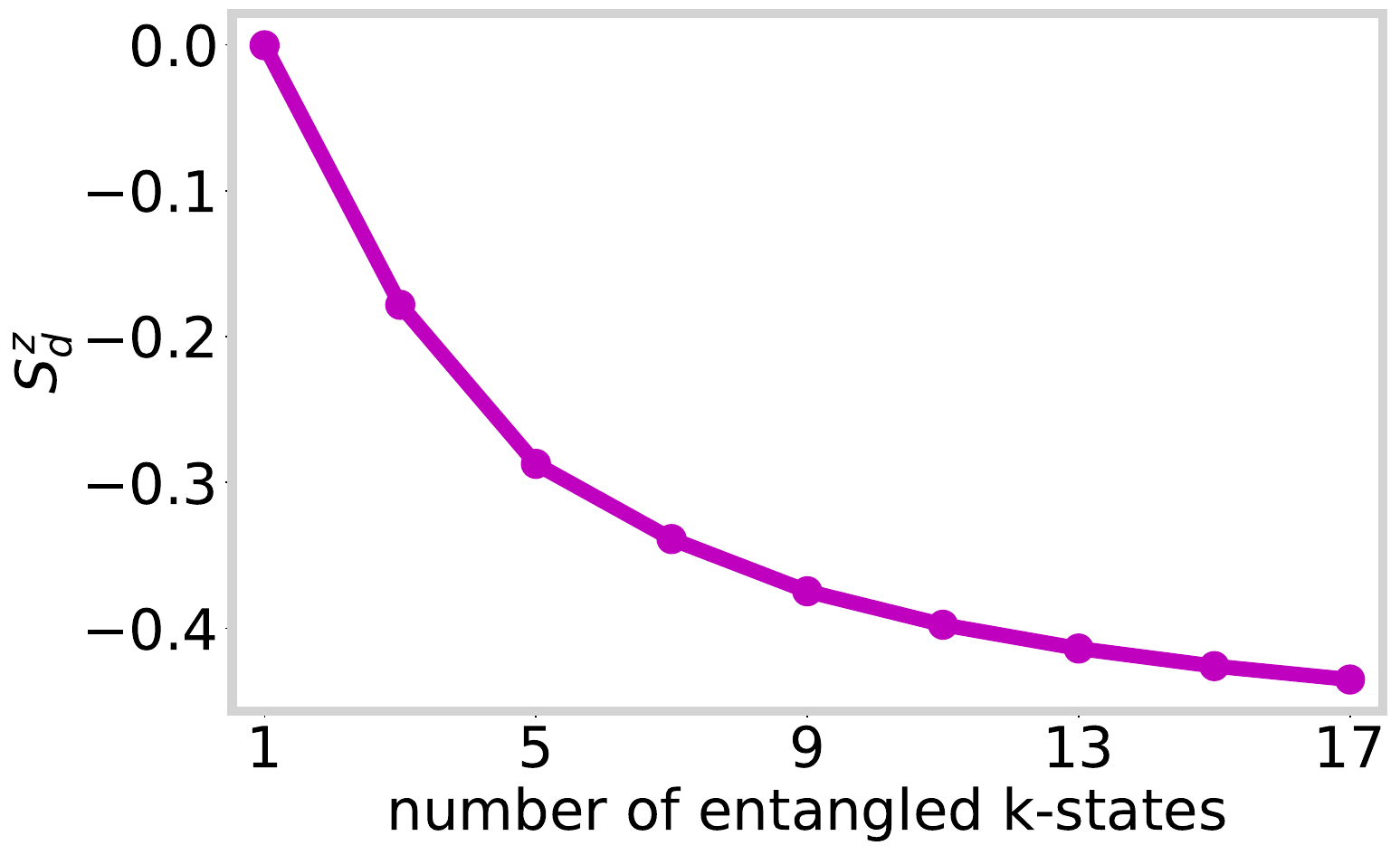}
    \includegraphics[scale=0.26]{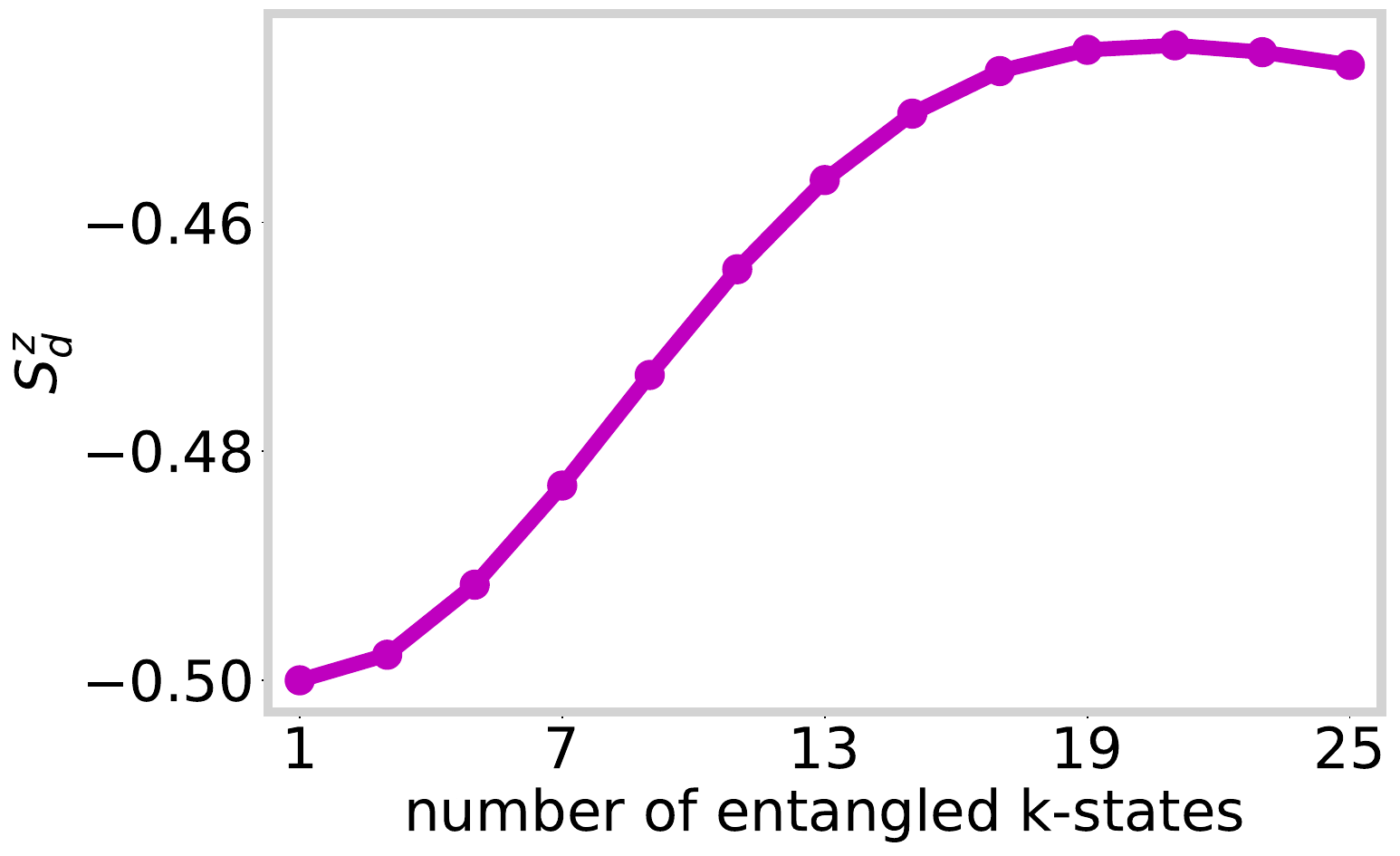}
    \caption{MERG evolution impurity magnetisation in (left) singlet regime and (right) local moment regime.}
    \label{O9p}
\end{figure}

Fig. \ref{O9p} shows the MERG evolution of the impurity spin magnetization $S_{d}^{z}$. In the singlet metallic regime (Fig. \ref{O9p}(a)), the impurity $S_{d}^{z}$ is zero in the IR and increases as new members are added at each step taken towards the UV; this indicates the emergence of Kondo screening in the IR. In contrast, the magnetization of the local moment phase is maximised under the flow from UV to IR (Fig. \ref{O9p}(b)). In this way, the MERG analysis clearly illustrates the differences between the RG flows for various quantities to the Kondo-screened (metal) and local moment (insulator).

\section{Conclusions and Outlook}\label{conclusionssection}
Our results reveal a measurement-driven entanglement phase transition in our modified Kondo problem, from a metallic singlet to a polarized state, controlled by a local  magnetic field. The transition has been thoroughly investigated using several methods, including the non-perturbative Unitary Renormalization Group (URG) and Momentum Space Entanglement Renormalization Group (MERG) formalisms. A deeper understanding of the metal-insulator transition is obtained by conducting a perturbation theoretic analysis of single particle hybridisation from the conduction bath on the low-energy fixed-point theory, obtaining insight on the non-Fermi liquid low-energy excitations emergent at the transition. This is further verified by computing various quantities, such as the impurity spectral function and quasiparticle residue. The transition is also analyzed via thermalization-related diagnostics and various impurity observables along the RG flow, relating them to the growth (or decay) of entanglement with the conduction bath under RG.  

Our work opens the door to several possible future directions. First, it would be interesting to conduct a similar study in which the purely deterministic continuous measurement (static local $B$-field) is replaced with a stochastic or time-dependent field (e.g., telegraph noise, periodic drive), and compare with randomness-induced transitions in monitored systems. This would help concretise the analogy to monitored circuits quantitative by extracting scaling forms (e.g., for impurity–bath entanglement measures and local magnetization) and identifying universal critical exponents across the transition. Second, the computation of additional observables (e.g., dynamic spin susceptibility, conductance for quantum-dot realizations of the Kondo problem) would help with proposing concrete protocols in experimental realisations such as quantum dots and STM studies Kondo systems. This would offer further insight into the field-tuned spectroscopy across the critical regime, as well as lead to engineered qubit-fermion bath simulators where the an external field plays the controlled measurement strength. Third, detailed comparisons with controlled experiments would also help make the connection between entanglement measures and measurables explicit, e.g., utilise thermalization diagnostics as a quantitative “order parameter” for transitions and crossover phenomena, and enable the experimental investigation of how well the phenomena of entanglement thermalization correlates with fixed-point structure and decoherence rates. Finally, we envisage extensions of our study at finite-temperature as well as under non-equilibrium situations (e.g., quenches/ramps of the magnetic field, Kibble-Zurek type scaling for adiabatic ramps across the critical regime).


\paragraph{Funding information}
S.L. thanks the SERB, Govt. of India for funding through MATRICS grant MTR/2021/000141 and Core Research Grant CRG/2021/000852. S.L. also thanks the Anusandhan National Research Foundation, Govt. of India for funding through Advanced Research Grant ANRF/ARG/2025/004414/PS. D.D. and A.M. thank IISER Kolkata for funding through JRF and SRF positions. 


\begin{appendix}
\section{Details of URG Calculation}
\label{A!w}
The Kondo Hamiltonian has a diagonal and an off-diagonal part. The diagonal part can be written as $\sum_q \Big( \varepsilon_q \tau_{q\sigma} + \frac{J}{2} S_d^z c_{q\uparrow}^\dagger c_{q\uparrow} - \frac{J}{2} S_d^z c_{q\downarrow}^\dagger c_{q\downarrow} + B \mu_B S_d^z \Big)$, where $\tau_{q\sigma} = n_{q\sigma} -\frac{1}{2}$. Similarly, the off-diagonal part can be written as
\begin{equation}
    H_1^I = \sum_{\abs{k}<\Lambda,q} \frac{J}{2} S_d^z c_{k\uparrow}^\dagger c_{q\uparrow} - \frac{J}{2} S_d^z c_{k\downarrow}^\dagger c_{q\downarrow} +\frac{1}{2} \sum_{\abs{k}<\Lambda,q} J\Big(S_d^+ S_{kq}^- + S_d^- S_{kq}^+ \Big) 
\end{equation}

and,
\begin{equation}
    H_0^I = \sum_{\abs{k}<\Lambda,q} \frac{J}{2} S_d^z c_{q\uparrow}^\dagger c_{k\uparrow} - \frac{J}{2} S_d^z c_{q\downarrow}^\dagger c_{k\downarrow} +\frac{1}{2} \sum_{\abs{k}<\Lambda,q} J\Big(S_d^+ S_{qk}^- + S_d^- S_{qk}^+ \Big) 
\end{equation}
 We consider scattering processes from high momentum $q$ to the Fermi surface. $\Lambda $ is maximum possible momentum due to bandwidth $D$.  The change in Hamiltonian at each RG step is given by
 \begin{align*}
 \Big(\Delta H_(j) \Big)_{\vec{q},\beta} = H_{(j-1)} - H_{(j)} = c_{\vec{q},\beta}^\dagger T_{\vec{q},\beta} \frac{1}{\omega - H_D} T_{\vec{q},\beta}^\dagger c_{\vec{q},\beta} + c_{\vec{q},\beta} \frac{1}{\omega - H_D} c_{\vec{q},\beta}^\dagger T_{\vec{q},\beta}
 \end{align*}
 \textbf{Calculation of RG equation for $J$:}\\
 Ising scattering, Particle sector:
 \begin{align*}
    & \sum_{\abs{k,k'}<\Lambda,q}[\frac{J}{2} S_d^z c_{q\uparrow}^\dagger c_{k'\uparrow} \frac{1}{w-H_{q\sigma}^D}  \frac{1}{2} J S_d^+S_{kq}^- - \frac{J}{2} S_d^z c_{q\downarrow}^\dagger c_{k'\downarrow} \frac{1}{w-H_{q\sigma}^D}  \frac{1}{2} J S_d^+S_{kq}^- \\ & \hspace{2cm} + \frac{J}{2} S_d^z c_{q\uparrow}^\dagger c_{k'\uparrow} \frac{1}{w-H_{q\sigma}^D}\frac{1}{2} J S_d^- S_{kq}^+  - \frac{J^{z}}{2} S_d^z c_{q\downarrow}^\dagger c_{k'\downarrow} \frac{1}{w-H_{q\sigma}^D}\frac{1}{2} J S_d^- S_{kq}^+] \\
    &= \sum_{\abs{k,k'}<\Lambda,q} [-\frac{1}{8} J^2  \frac{1}{w-E_{11}} S_d^+ S_{kk'}^-  n_{q\uparrow}  - \frac{1}{8} J^2 \frac{1}{w-E_{12}} S_d^- S_{kk'}^+  n_{q\downarrow} ]
\end{align*}

$ \frac{1}{w-H_{q\sigma}^D} = \frac{1}{w-E_{11}} $ where $E_{11}= \frac{D}{2} -\frac{J^{z}}{4} +\frac{\mu_B B}{2}$ when $S_d^z= \frac{1}{2}$ and $q\downarrow$ is present. Also, $S_d^z= -\frac{1}{2}$ and $q\uparrow$ gives 
$ \frac{1}{w-H_{q\sigma}^D} = \frac{1}{w-E_{12}} $ where $E_{12}= \frac{D}{2} -\frac{J^{z}}{4} -\frac{\mu_B B}{2}$.\\
Spin-flip scattering, Particle sector:
\begin{align*}
    & \sum_{\abs{k,k'}<\Lambda,q} \Big[\frac{J}{2} S_d^+ S_{q k'}^- \frac{1}{w-H_{q\sigma}^D}\frac{J}{2} S_d^z c_{k\uparrow}^\dagger c_{q\uparrow}  -\frac{J}{2}S_d^+ S_{q k'}^- \frac{1}{w-H_{q\sigma}^D} \frac{J}{2} S_d^z c_{k\downarrow}^\dagger c_{q\downarrow} \\  & \hspace{1.5cm} + \frac{J}{2} S_d^- S_{q k'}^+ \frac{1}{w-H_{q\sigma}^D}\frac{J}{2} S_d^z c_{k\uparrow}^\dagger c_{q\uparrow} -\frac{J}{2} S_d^- S_{q k'}^+ \frac{1}{w-H_{q\sigma}^D} \frac{J}{2} S_d^z c_{k\downarrow}^\dagger c_{q\downarrow} \\
     & = \sum_{\abs{k,k'}<\Lambda,q} \Big[- \frac{J^2}{8} \frac{1}{w-E_{12}} S_d^+  S_{kk'}^- n_{q\downarrow}  - \frac{J^2}{8} \frac{1}{w-E_{11}} S_d^- S_{k k'}^+ n_{q\uparrow} \\ & \hspace{2.5cm} - \frac{J^2}{4} \frac{1}{w-E_{12}} S_d^z c_{k \uparrow}^\dagger c_{k'\uparrow} n_{q\downarrow}  + \frac{J^2}{4} \frac{1}{w-E_{11}} S_d^z c_{k \downarrow}^\dagger c_{k'\downarrow} n_{q\uparrow}  \Big] 
\end{align*}

Total RG equation in Particle sector:
\begin{align*}
\sum_{\abs{k,k'}<\Lambda,q} [-\frac{J^2}{8} \frac{1}{w-E_{11}} (S_d^+ S_{kk'}^-  + S_d^- S_{kk'}^+ )  -\frac{J^2}{8} \frac{1}{w-E_{12}} ( S_d^+ S_{kk'}^- + S_d^- S_{kk'}^+ )\\
    - \frac{J^2}{4}   \frac{1}{w-E_{12}} S_d^z c_{k \uparrow}^\dagger c_{k'\uparrow}  + \frac{J^2}{4} \frac{1}{w-E_{11}} S_d^z c_{k \downarrow}^\dagger c_{k'\downarrow} ]N(D) \Delta D 
\end{align*}
where $\sum_q n_{q\uparrow}=N(D) \Delta D$, $\sum_q n_{q\downarrow}=N(D) \Delta D$, and $N(D)$ is density of states of up-spin or down-spin.\\
Similarly, for the hole sector, we take $\sum_q (1-n_{q\uparrow})=N(D) \Delta D $ and $\sum_q (1-n_{q\downarrow})=N(D) \Delta D$. The RG equation for the hole sector is:
\begin{align*}
\sum_{\abs{k,k'}<\Lambda,q} [-\frac{J^2}{8} \frac{1}{w-E_{11}} (S_d^+ S_{k'k}^- + S_d^- S_{k'k}^+ )  -\frac{J^2}{8} \frac{1}{w-E_{12}} (S_d^+ S_{k'k}^- + S_d^- S_{k'k}^+ ) \\ - \frac{J^2}{4}  \frac{1}{w-E_{11}} S_d^z c_{k' \uparrow}^\dagger c_{k\uparrow}  + \frac{J^2}{4} \frac{1}{w-E_{12}} S_d^z c_{k' \downarrow}^\dagger c_{k\downarrow} ] N(D) \Delta D 
\end{align*}
The difference between the particle and hole sectors arising from the magnetic field can be seen in the second line of both the equations, giving the RG equation for $J$ as
\begin{equation}
    \frac{\Delta J}{\Delta D} = - \frac{J^2}{2} [\frac{ 1 }{w-E_{11}} +  \frac{ 1 }{w-E_{12}} ]N(D) ~.
\end{equation}

\textbf{Calculation of RG equation of $B$:}
Particle sector:
\begin{align*}
    \sum_{\lambda}
    \frac{1}{2} J^t S_d^+ S_{k_F +\lambda, k_F -\lambda}^- \frac{1}{w - H_{q\sigma}^D} \frac{1}{2} J^t S_d^- S_{k_F -\lambda, k_F +\lambda}^+  = \frac{1}{4} (J^t)^2 \frac{1}{w - E_{12}} S_d^z N(D) \Delta D~,
\end{align*}
and Hole sector:
\begin{align*}
    \sum_{\lambda}
    \frac{1}{2} J^t S_d^- S_{k_F +\lambda, k_F -\lambda}^+ \frac{1}{w - H_{q\sigma}^D} \frac{1}{2} J^t S_d^+ S_{k_F -\lambda, k_F +\lambda}^-  = - \frac{1}{4} (J^t)^2 \frac{1}{w - E_{11}} S_d^z N(D) \Delta D~,
\end{align*}
yield the RG equation of $B$ as
\begin{equation*}
    \mu_B \frac{\Delta B}{\Delta D} = - \frac{J^2}{4} [\frac{1}{w - E_{11}} -\frac{1}{w - E_{12}} ] N(D)~.
\end{equation*}
\section{Thermodynamic limit}
\label{S@3}
\begin{figure}[!ht]
    \centering
    \includegraphics[scale=0.5]{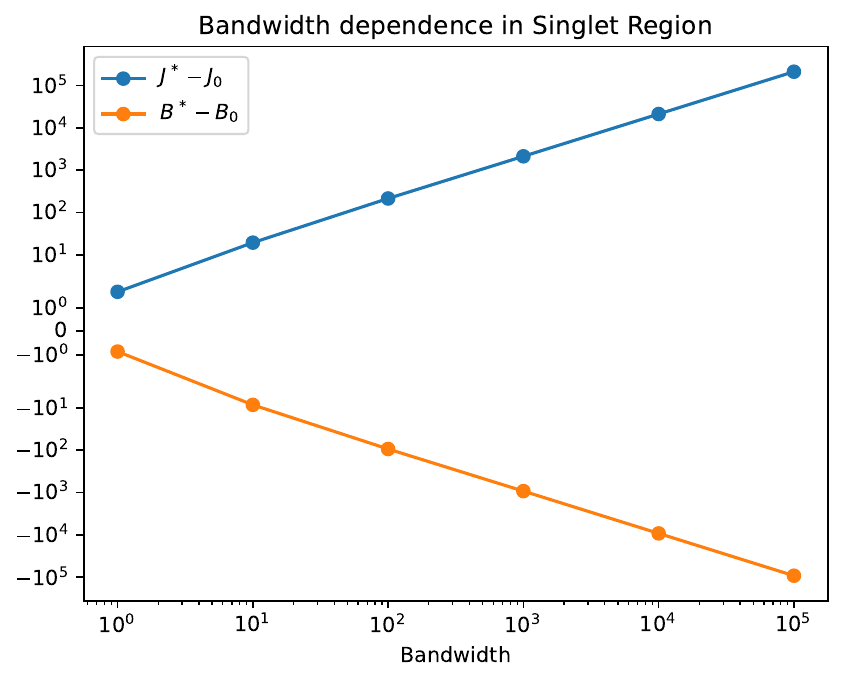}
    \includegraphics[scale=0.5]{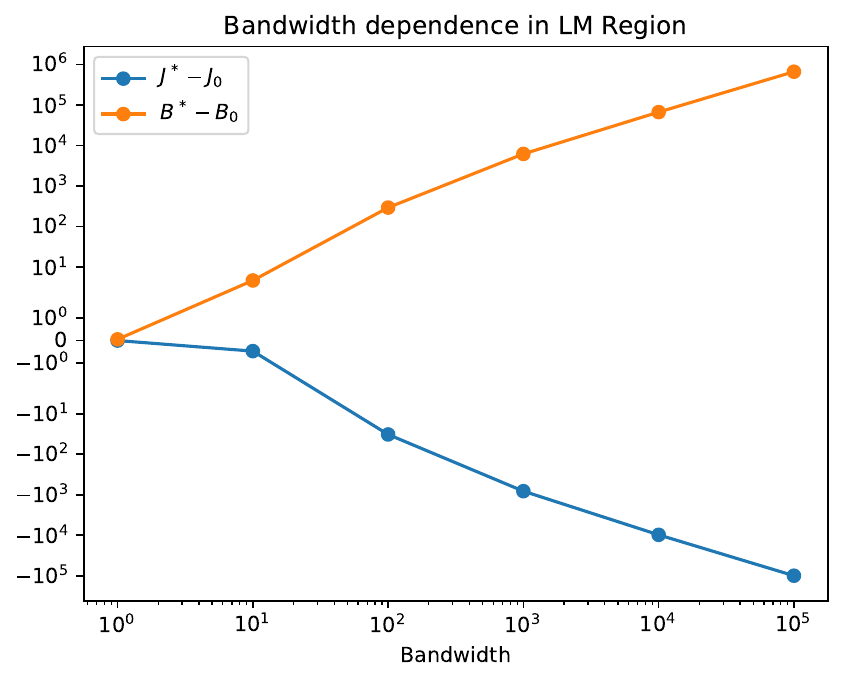}
    \caption{Bandwidth dependence of fixed-point Kondo coupling in (left) singlet regime and (right) local moment regime.}
    \label{TLB}
\end{figure}

Increasing the bandwidth reveals the following: in the metallic regime, the Kondo coupling constant \( J \) increases while the magnetic field \( B \) decreases. Conversely, the opposite behavior is seen in the local moment regime. This behavior is depicted in Figure \ref{TLB}.

\section{Results for the zero-bandwidth model}
\label{WH3}
The zero-bandwidth approximation involves neglecting the effects of single particle hybridisation with the the conduction bath, thereby simplifying the problem to a purely quantum mechanical one. As seen in Section \ref{CritPeffH}, a perturbation theoretic treatment of the electron hopping into the conduction bath from the ground states of the zero-bandwidth Hamiltonian allows us to gain insights into the behavior of the low-energy excitations of the full fixed-point Hamiltonian. 

The zero-bandwidth Hamiltonian, expressed in dimensionless form, is given by:
\begin{align*}
H' = \frac{H}{J} = S_d \cdot S_0 + g S_d^z~,
\end{align*}
where $g=\frac{B^* \mu_B}{J^*}$.

\begin{figure}[!ht]
\centering
\includegraphics[scale=0.45]{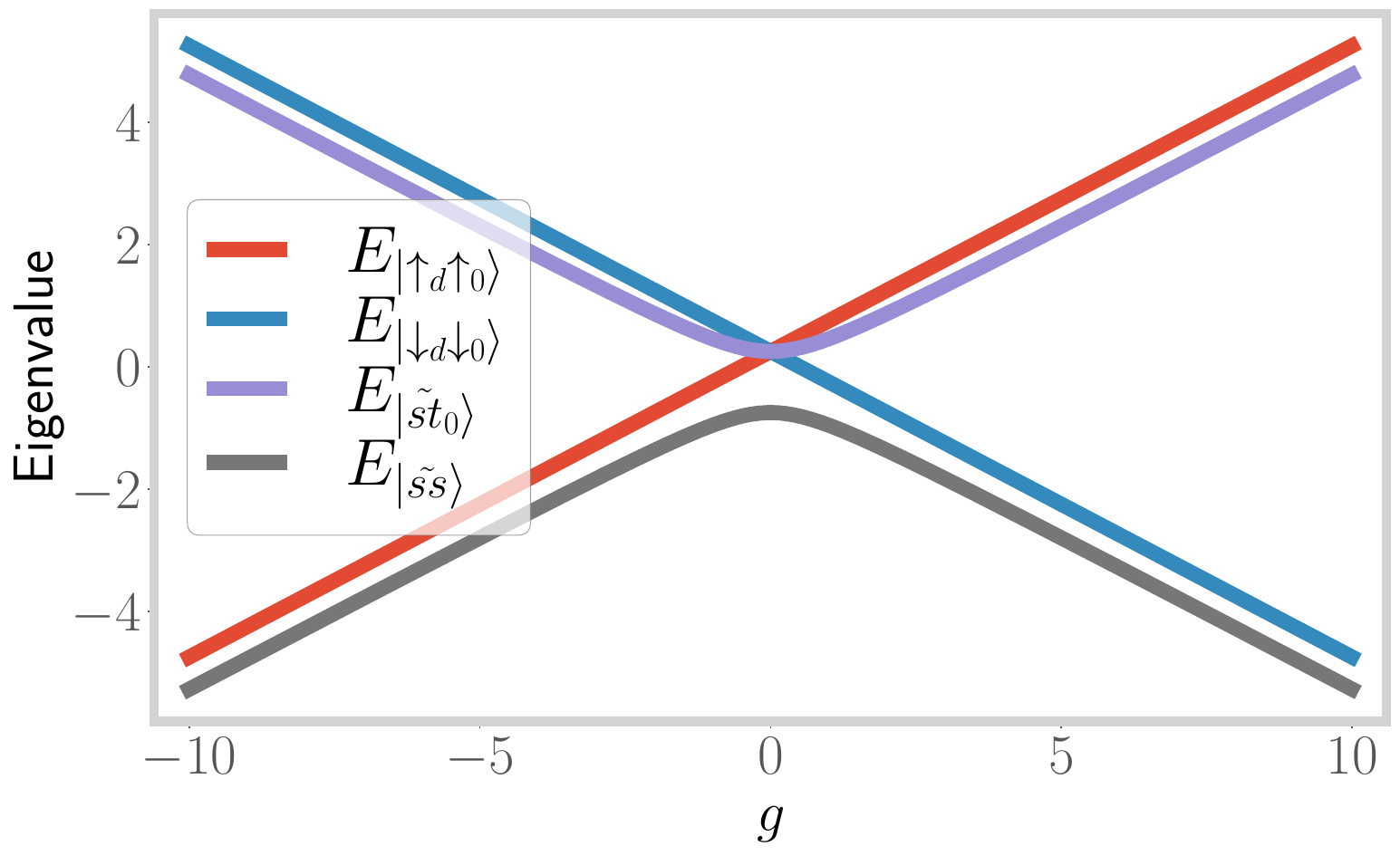}
\caption{Variation of eigenvalues of zero-bandwidth model with the dimensionless parameter $g=\frac{\mu_B B}{J}$.}
\label{oo}
\end{figure}

The eigenstates and corresponding eigenvalues are:

\begin{center}
	\begin{tabular}{ |c | c | }
  	\hline			
  	Eigenstate & Eigenvalue \\
  	\hline
  	$\ket{\uparrow_d \uparrow_0}$ & $\frac{1}{4} + \frac{g}{2}$  \\
  	\hline
 	 $\ket{\Tilde{st}_0} = \frac{1}{\sqrt{2}} \frac{1}{\sqrt{1+g^2+g\sqrt{1+g^2}}} \Big[(g+\sqrt{1+g^2}) 		\ket{\uparrow_d \downarrow_0} +\ket{\downarrow_d \uparrow_0} \Big]$  & $-\frac{1}{4} + \frac{\sqrt{1 + g^2}}{2}$ \\
  	\hline  
  	$\ket{\Tilde{ss}} = - \frac{1}{\sqrt{2}} \frac{1}{\sqrt{1+g^2+g\sqrt{1+g^2}}} \Big[\ket{\uparrow_d \downarrow_0} -(g+\sqrt{1+g^2}) \ket{\downarrow_d \uparrow_0} \Big]$ & $-\frac{1}{4} - \frac{\sqrt{1 + g^2}}{2}$ \\
  	\hline
  	$\ket{\downarrow_d \downarrow_0} $ & $\frac{1}{4} -\frac{g}{2}$ \\
  	\hline
	\end{tabular}
\end{center}

Here, $\ket{\Tilde{st}_0}$ and $\ket{\Tilde{ss}}$ represent the distorted spin-zero triplet and distorted spin singlet, respectively. For $g=0$, these states reduce to the familiar spin-zero triplet and singlet states. As $g$ approaches infinity, $\ket{\Tilde{ss}}$ transforms into $ \ket{\downarrow_d \uparrow_0}$. Figure \ref{oo} illustrates this behavior, showing that the eigenvalues $E_{\ket{\Tilde{ss}}}$ and $E_{\ket{\downarrow_d \downarrow_0}}$ become degenerate in the asymptotic limit of $g\to\infty$. Importantly, in Section \ref{CritPeffH}, we use the degeneracy of these two states at a finite value of the dimensionless coupling $g^{*}$ (obtained from the RG equations) to compute the effective Hamiltonian for the low-energy non-Fermi liquid equations.

\subsection{Magnetisation for localized d-electron}
\begin{figure}[!ht]
\centering
\includegraphics[scale=0.45]{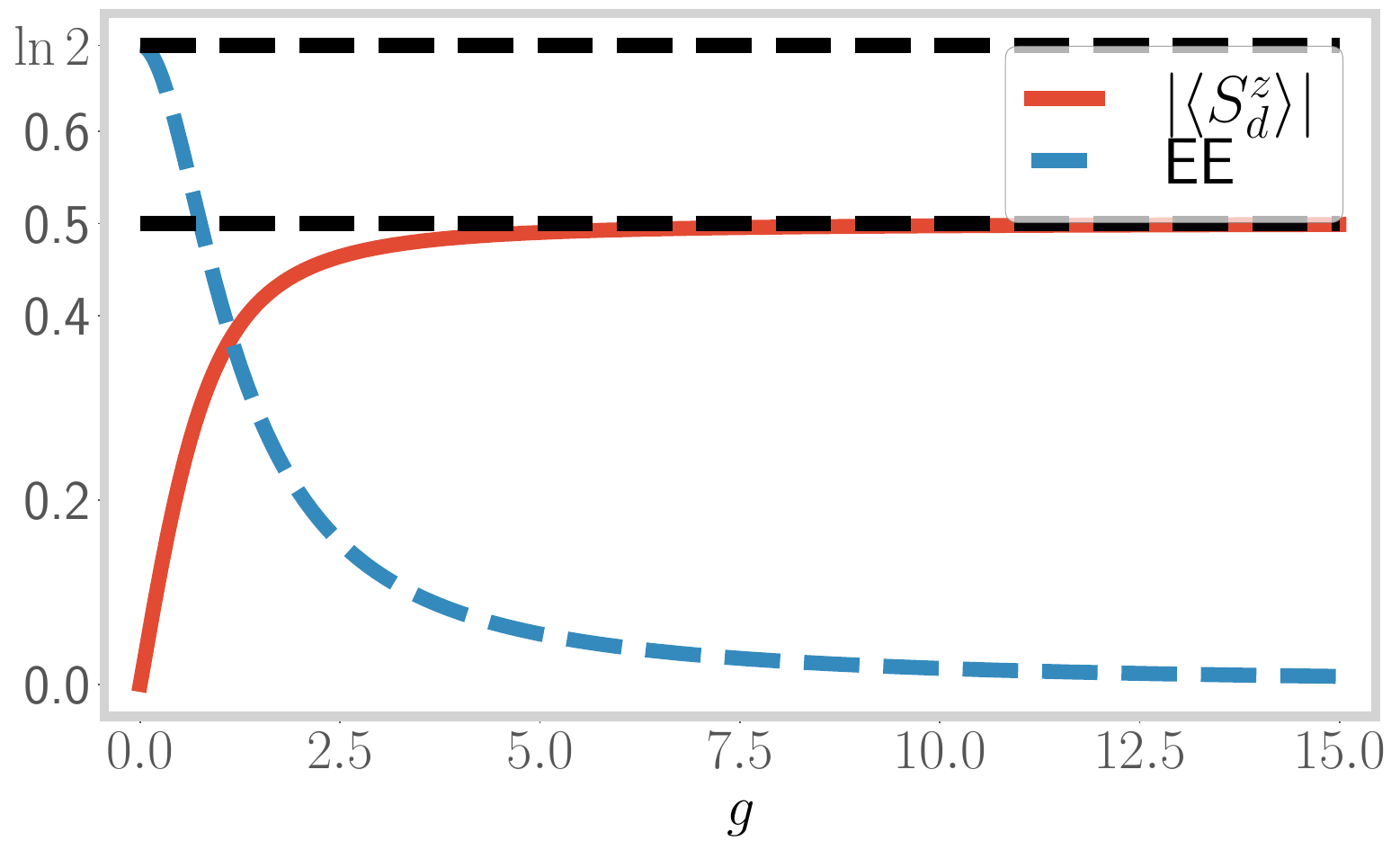}
\caption{Variation of impurity magnetization ($S_{d}^{z}$) and impurity entanglement entropy (EE) with $g=\frac{\mu_B B}{J}$.}
\label{o1}
\end{figure}

In the absence of an applied magnetic field, the magnetization of the d-electron for the ground state $\ket{\Tilde{ss}}$ is zero. We now investigate the magnetization induced by an applied magnetic field. The magnetization is given by:
\begin{align*}
&\langle S_d^z \rangle  =  -\frac{1}{2}\frac{g^2 +g \sqrt{1+g^2}}{1+ g^2+ g\sqrt{1 + g^2}} 
\end{align*} 
As the magnetic field attempts to polarize the impurity spin downward, a large B-field ($g >> 1$) results in the magnetization $ | \langle S_d^z \rangle | $ approaching $\frac{1}{2}$. This behavior is depicted in Figure \ref{o1}.

\subsection{Entanglement Entropy}
The ground state singlet Entanglement Entropy (EE) for the Kondo model is $\ln 2$. We now discuss the impurity EE ($S_d (EE)$)in the presence of a magnetic field. The reduced density matrix of the impurity site, obtained by tracing out the degrees of freedom on the zeroth site of the zero-bandwidth model, is given by:
\begin{align*}
\rho_d 
& =\begin{bmatrix}
\frac{1}{2} + \langle S_d^z \rangle & 0 \\
0 & \frac{1}{2} - \langle S_d^z \rangle
\end{bmatrix}
\end{align*}
Using the expression for $ \langle S_d^z \rangle $ in terms of $g$, the entanglement entropy(EE) becomes:
\begin{align*}
S_d (EE) & =- Tr[\rho_d \ln \rho_d] \\
& = - \frac{1}{2}\frac{1}{1+ g^2+ g\sqrt{1 + g^2}} \ln \Big( \frac{1}{2}\frac{1}{1+ g^2+ g\sqrt{1 + g^2}} \Big) \\ & \hspace{.5cm} - \frac{1}{2}\frac{ \{g + \sqrt{1 + g^2} \}^2}{1+ g^2+ g\sqrt{1 + g^2}} \ln \Big( \frac{1}{2}\frac{ \{g + \sqrt{1 + g^2} \}^2}{1+ g^2+ g\sqrt{1 + g^2}} \Big)
\end{align*}
As the magnetic field increases, the impurity spin increasingly aligns with the field. In the limit of a large B-field ($g >> 1$), the impurity spin is fully polarized, leading to the destruction of the entangled state. Consequently, the EE approaches zero. Figure \ref{o1} illustrates this behavior, showing that as $g$ increases, the EE decreases, reflecting the loss of entanglement as the system transitions to a polarized state. 

\subsection{Entanglement dynamics of the ground state}
We first choose an initial state  
\begin{align*}
\ket{\downarrow_d \uparrow_0} = \frac{1}{\sqrt{2}} \frac{1}{\sqrt{1+g^2+g\sqrt{1+g^2}}} \Big[\ket{\Tilde{st}_0} + (g+\sqrt{1+g^2}) \ket{\Tilde{ss}} \Big]~.
\end{align*}
The time-dependent form of this state can be written as :
\begin{align}
\ket{\downarrow_d \uparrow_0}(t) = \frac{1}{\sqrt{2}} \frac{1}{\sqrt{1+g^2+g\sqrt{1+g^2}}} \Big[e^{-\frac{i}{\hbar}E_{\Tilde{st}_0} t} \ket{\Tilde{st}_0} + (g+\sqrt{1+g^2}) e^{-\frac{i}{\hbar}E_{\Tilde{ss}} t} \ket{\Tilde{ss}} \Big]
\end{align}

The probabilities of obtaining the distorted singlet and distorted spin zero triplet states from the time-dependent state $\ket{\downarrow_d \uparrow_0}_{(t)}$ are given by:
\begin{align}
P_{\bra{\Tilde{ss}} \ket{\downarrow_d \uparrow_0}}(t) =\frac{1}{2} -\langle S_d^z \rangle \equiv P_y \hspace{1cm}~~,~~P_{\bra{\Tilde{st}_0} \ket{\downarrow_d \uparrow_0}} (t)=\frac{1}{2} +\langle S_d^z \rangle \equiv P_x~.
\end{align}

\begin{figure}[!ht]
    \centering
    \includegraphics[scale=0.25]{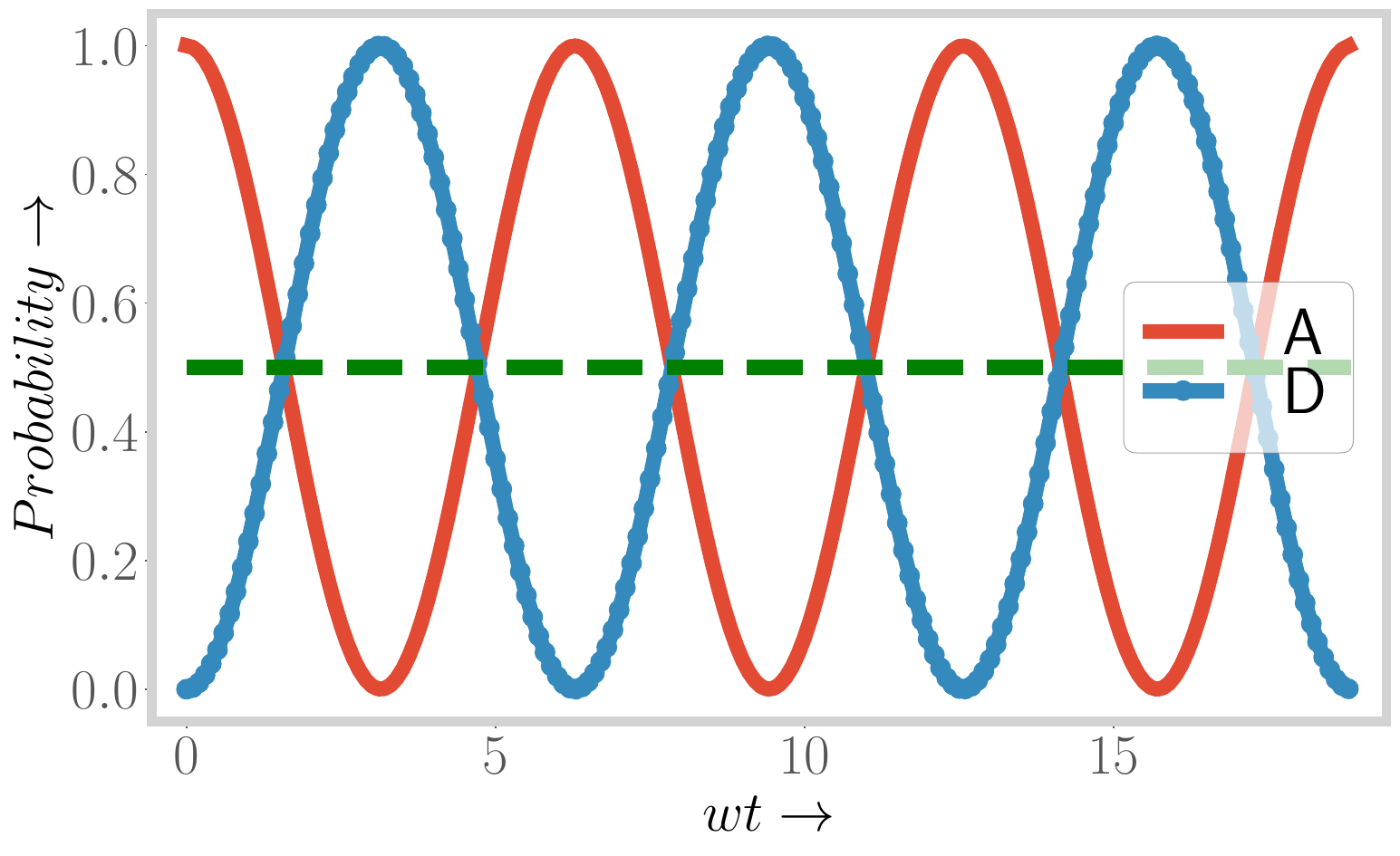}
    \includegraphics[scale=0.25]{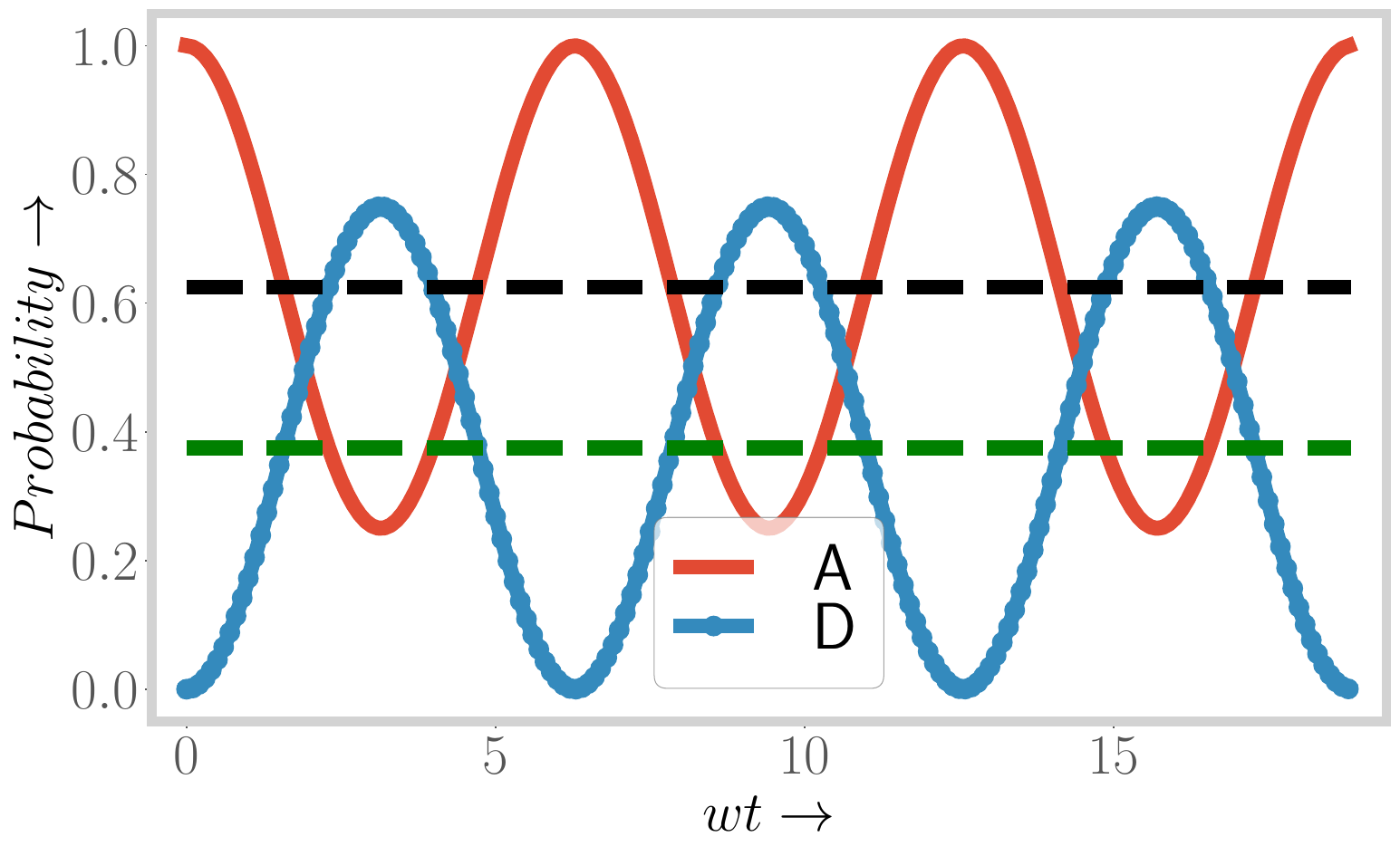}
    \includegraphics[scale=0.25]{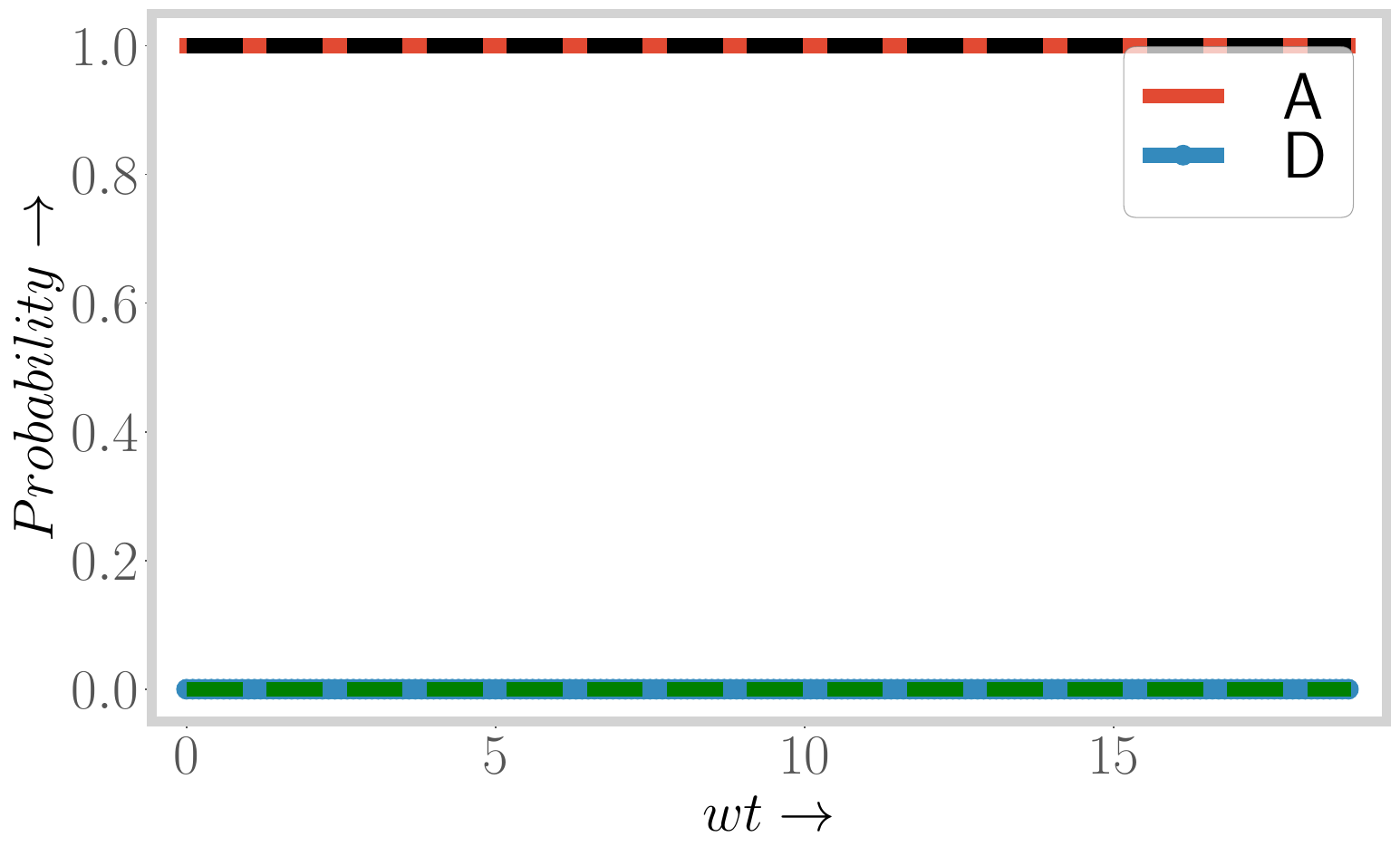}
    \caption{Time evolution of probability coefficients $A$ and $D$ (see text for details) for several values of $\langle S_d^z \rangle$: \textbf{Upper left} $\langle S_d^z \rangle =0 $, \textbf{upper right} $\langle S_d^z \rangle = 0.25$ and \textbf{below} $\langle S_d^z \rangle=0.5$.}
    \label{o091}
\end{figure}

Time independent classical $\ket{\downarrow_d \uparrow_0}_{(0)}$ and $\ket{\uparrow_d \downarrow_0}_{(0)}$ can be found in time dependent classical $\ket{\downarrow_d \uparrow_0}_{(t)}$ as a probability :
\begin{align}
P_{\bra{\downarrow_d \uparrow_0}_{(0)} \ket{\ket{\downarrow_d \uparrow_0}}}(t) = P_1 +4 \langle S_d^z \rangle^2 P_2~~,~~
P_{\bra{\uparrow_d \downarrow_0}_{(0)} \ket{\ket{\downarrow_d \uparrow_0}}}(t)  = (1 - 4 \langle S_d^z \rangle^2 ) P_2~,
\end{align}
where $P_1 = \frac{1}{2} \left( 1+ \cos \left(\frac{E_{\Tilde{st}_0} - E_{\Tilde{ss}}}{\hbar}t \right) \right)$ and $ P_2 = \frac{1}{2} \left( 1 - \cos \left(\frac{E_{\Tilde{st}_0} - E_{\Tilde{ss}}}{\hbar}t \right) \right)$~,
such that
\begin{align*}
P_{\bra{\downarrow_d \uparrow_0}_{(0)} \ket{\ket{\downarrow_d \uparrow_0}_{(t)}}} + P_{\bra{\uparrow_d \downarrow_0}_{(0)} \ket{\ket{\downarrow_d \uparrow_0}_{(t)}}} =1~.
\end{align*}
Similar results can also be obtained for an initial state of $\uparrow_d \downarrow_0$.\\
The time evolution of reduced density matrix $\rho_d(t)$ for $\ket{\uparrow_d \downarrow_0}_{(t)}$ gives
\begin{align*}
\ket{\uparrow_d \downarrow_0}(t)
& = \frac{1}{2} \frac{1}{1+g^2+g\sqrt{1+g^2}} \Big[\Big((g+\sqrt{1+g^2})^2 e^{-\frac{i}{\hbar}E_{\Tilde{st}_0} t} + e^{-\frac{i}{\hbar}E_{\Tilde{ss}}t} \Big) \ket{\uparrow_d \downarrow_0} \\& \hspace{2.5cm} + (g+\sqrt{1+g^2}) \Big(e^{-\frac{i}{\hbar}E_{\Tilde{st}_0} t} - e^{-\frac{i}{\hbar}E_{\Tilde{ss}}t} \Big) \ket{\downarrow_d \uparrow_0} \Big]~, 
\end{align*}
and the time-evolved reduced density matrix as:
\begin{align*}
\rho^d(t)
& = A \ket{\uparrow_d} \bra{\uparrow_d} + D \ket{\downarrow_d} \bra{\downarrow_d}~.
\end{align*}
Thus, we find that
\begin{align*}
\begin{pmatrix}
1 & 0\\
0 & 0
\end{pmatrix} \underrightarrow{\rho^d\text{ time dynamics}} \begin{pmatrix}
A & 0\\
0 &  D
\end{pmatrix}~,
\end{align*}
where $A  = P_1 + 4 \langle S_d^z \rangle^2  P_2 = 1 - 2 P_x P_y \Biggl(1 - \cos \left(\frac{E_{\Tilde{st}_0} - E_{\Tilde{ss}}}{\hbar}t \right) \Biggl)$ and \\$D = (1 - 4 \langle S_d^z \rangle^2 ) P_2 = 2P_x P_y \Biggl( 1- \cos \left(\frac{E_{\Tilde{st}_0} - E_{\Tilde{ss}}}{\hbar}t \right) \Biggl) $~.\\
In Fig.\ref{o091}, we present the time evolution of the coefficients $A$ and $D$ for several values of $\langle S_d^z \rangle$. 

\section{Coefficients of effective Hamiltonian at 2nd order}
\label{W13}
The coefficients of the effective Hamiltonian at second order are given by 
\begin{eqnarray}
\mathcal{J}^\perp &=& 2 J C_4~,~\mathcal{J}^z=2 J (C_3-C_2-C_1)~,~\mathcal{H}_0^1 = J(C_1+C_3-C_2)~,~\mathcal{H}_0^2=J(C_2+C_3-C_1)~,~\textrm{where}\nonumber\\
C_1 &=&\frac{4t^2}{J^2(2g-1-2\sqrt{1+g^2})}~,~C_2=\frac{4t^2 }{J^2(2g -1 - 2\sqrt{1+g^2})}\frac{1}{2}\frac{(g+ \sqrt{1 + g^2})^2}{1+ g^2+ g\sqrt{1 + g^2}}~,\nonumber\\
C_3 &=&-\frac{4t^2 }{J^2(2g +1 + 2\sqrt{1+g^2})}\frac{1}{2}\frac{1}{1+ g^2+ g\sqrt{1 + g^2}}~,\nonumber\\
C_4 &=&-\frac{4t^2}{J^2(2g-1-2\sqrt{1+g^2})} \frac{1}{\sqrt{2}}\frac{g + \sqrt{1 + g^2}}{\sqrt{1+ g^2+ g\sqrt{1 + g^2}}}~.\nonumber   
\end{eqnarray}

\section{Coefficients of effective Hamiltonian at 4th order}
\label{D&5}
The coefficients of the effective Hamiltonian at fourth order are given by
\begin{eqnarray}
\alpha &=& Y_3 + (N_1-Y_3)~,~\beta = Y_3~, ~\gamma =  Y_5 + (Y_4 -Y_5)~,\nonumber\\ 
\eta &=& ( N_2 - Y_5 )~,~\zeta = A_0^2 b_1 \Big( Y_1 (\lambda_+ - a_1) + Y_2 (\lambda_- - a_1) \Big)~,~\textrm{where}\nonumber\\ 
a_1&=&\frac{2}{E_i +\frac{g}{2}}\frac{t^2}{J^2}
~,~b_1= \frac{2 C_A C_B} {E_i +\frac{g}{2}} \frac{t^2}{J^2}~,~f_1 =\frac{2 C_A^2 C_B^2}{E_i + \frac{g}{2}} \frac{t^2}{J^2}~,~d_1=\frac{2 C_A}{E_i + \frac{g}{2}} \frac{t^2}{J^2}~,~d_2=\frac{2 C_A^2 C_B} {E_i + \frac{g}{2}} \frac{t^2}{J^2}~,\nonumber\\
C_A &=&-\frac{1}{\sqrt{2}}\frac{1}{\sqrt{1+ g^2+ g\sqrt{1 + g^2}}}~,~C_B=g + \sqrt{1 + g^2}~,~A_0=\frac{1}{\sqrt{b_1^2 + (\lambda_{\pm} - a_1)^2}}~,\nonumber\\
N_1 &=& A_0^2 b_1^2 (Y_1 + Y_2)~,~ N_2 = A_0^2 Y_1 (\lambda_+ - a_1)^2 + A_0^2 Y_2 (\lambda_- - a_1)^2~,\nonumber\\
Y_{1}&=&-\frac{A_0^2 (b_1 d_1 + \lambda_{+} d_2 - a_1 d_2 )^2}{\sqrt{1+g^2}}\nonumber\\ 
&&- A_0^4 \Big( \lambda_+ b_1^2 + ( \lambda_+ - a_1 ) ( b_1^2 + \lambda_+ f_1 -a_1 f_1 ) \Big) \cross \Big(\frac{a_1^2 b_1^2}{2} + a_1 b_1^2 (\lambda_+ - a_1)+ (\lambda_+ - a_1)^2 \frac{a_1 f_1}{2} \Big)~,\nonumber\\      
Y_2 &=&-\frac{A_0^2 (b_1 d_1 + \lambda_{-} d_2 - a_1 d_2 )^2}{\sqrt{1+g^2}}\nonumber\\ 
&&- A_0^4 \Big( \lambda_- b_1^2 + ( \lambda_- - a_1 ) ( b_1^2 + \lambda_- f_1 -a_1 f_1 ) \Big)  \cross \Big(\frac{a_1^2 b_1^2}{2} + a_1 b_1^2 (\lambda_- - a_1)+ (\lambda_- - a_1)^2 \frac{a_1 f_1}{2} \Big)~,\nonumber\\
Y_{3}&=&- \frac{1}{(E_i + \frac{g}{2})^3} \frac{t^4}{J^4} ~,~\lambda_{\pm} = \frac{(a_1 +f_1) \pm \sqrt{(a_1 - f_1)^2 + 4b_1^2}}{2}~,\nonumber\\
Y_4&=&-\frac{C_A^4 C_B^2}{\sqrt{1+g^2}} \frac{t^4}{J^4} \Big( \frac{1} {E_i - \frac{g}{2}} - \frac{1} {E_i + \frac{g}{2}} \Big)^2 -  C_A^4 \frac{t^4}{J^4}  \Big(\frac{1}{E_i - \frac{g}{2}} + \frac{C_B^2} {E_i + \frac{g}{2}} \Big) \Big(\frac{1}{(E_i - \frac{g}{2})^2} + \frac{C_B^2} {(E_i + \frac{g}{2})^2} \Big)~,\nonumber\\
Y_{5}&=&\frac{4 C_A^2}{(E_i - \frac{g}{2})^2} \frac{t^4}{J^4} \Big(- \frac{C_A^2 C_B^2} {\sqrt{1+g^2}} + \frac{1} {E_i - (\frac{1}{4} +\frac{g}{2})} \Big) - \frac{4 C_A^4}{(E_i -\frac{g}{2})^3} \frac{t^4}{J^4}~.
\end{eqnarray}

\section{Analytical expression for Kondo Temperature}
\label{D4E}
We first derive the Kondo coupling($J$) the from magnetic field-induced single impurity Anderson model through a Schrieffer-Wolf transformation~\cite{schrieffer1966}. We then insert the magnetic field-dependent $J$ in the equation of the Kondo temperature $T_k = D_0 \sqrt{2J\rho} \hspace{1mm} e^{-\frac{1}{2J\rho}}$.\\
From Schrieffer-Wolf transformation~\cite{schrieffer1966relation}, we obtain the change in the Hamiltonian as
\begin{align*}
	\Delta H = \frac{1}{2} \sum_{H}\Big[V\frac{\ket{H}\bra{H}}{E - E_H} V + V\frac{\ket{H}\bra{H}}{E' - E_H} V \Big]~, 
\end{align*}
Where $H$ is the high-energy subspace, $E$ is the energy of a state within the low-energy subspace and $E'$ is the energy of a state within the high energy subspace.
\begin{figure}[!ht]
    \centering
    \includegraphics[scale=0.85]{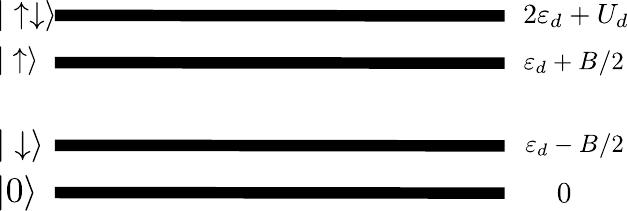}
    \caption{Schematic diagram of energy level of the Anderson impurity.}
\end{figure}
For a particle-hole symmetric case($\varepsilon_d = - \frac{U_d}{2}$) and $U_d >> B/2$, the high-energy subspace includes the states $\ket{\uparrow \downarrow}$ and $\ket{0}$ (at energy $E=0$). The low-energy subspace includes the states $ \ket{\uparrow}$ (with energy $E=- \frac{U_d}{2} + \frac{B}{2}$) and $ \ket{\downarrow}$ (with energy $E=- \frac{U_d}{2} - \frac{B}{2}$). For $E = E'$, the change in the Hamiltonian is given by
\begin{align*}
	\Delta H &= \sum_{H} V\frac{\ket{H}\bra{H}}{E - E_H} V~,\\ 
	& = \sum_{k_1,K_2} \Big[ \frac{V c_{d\uparrow}^\dagger c_{k_2 \uparrow} \ket{0} \bra{0} V c_{k_1\uparrow}^\dagger c_{d\uparrow}}{\frac{B}{2} -\frac{U_d}{2}} + \frac{V c_{d\downarrow}^\dagger c_{k_2\downarrow}\ket{0} \bra{0} V c_{k_1\downarrow}^\dagger c_{d\downarrow}}{-\frac{B}{2} -\frac{U_d}{2}} + \frac{V c_{k_1\uparrow}^\dagger c_{d\uparrow} \ket{\uparrow \downarrow} \bra{\uparrow \downarrow} V c_{d\uparrow}^\dagger c_{k_2 \uparrow}}{\frac{B}{2} -\frac{U_d}{2}} \\
	&+ \frac{V c_{k_1\downarrow}^\dagger c_{d\downarrow} \ket{\uparrow \downarrow} \bra{\uparrow \downarrow} V c_{d\downarrow}^\dagger c_{k_2\downarrow} }{-\frac{B}{2} -\frac{U_d}{2}}  + \frac{1}{2} V c_{d\uparrow}^\dagger c_{k_2 \uparrow} \ket{0} \bra{0} V c_{k_1\downarrow}^\dagger c_{d\downarrow} \Big(\frac{1}{\frac{B}{2} -\frac{U_d}{2}} + \frac{1}{-\frac{B}{2} -\frac{U_d}{2}} \Big) \\
	&+ \frac{1}{2} V c_{d\downarrow}^\dagger c_{k_2 \downarrow} \ket{0} \bra{0} V c_{k_1\uparrow}^\dagger c_{d\uparrow} \Big(\frac{1}{\frac{B}{2} -\frac{U_d}{2}} + \frac{1}{-\frac{B}{2} -\frac{U_d}{2}} \Big)\\  
    &+ \frac{1}{2} V c_{k_1\uparrow}^\dagger c_{d \uparrow} \ket{\uparrow \downarrow} \bra{\uparrow \downarrow} V c_{d \downarrow}^\dagger c_{k_2 \downarrow} \Big(\frac{1}{\frac{B}{2} -\frac{U_d}{2}} + \frac{1}{-\frac{B}{2} -\frac{U_d}{2}} \Big)  \\
	& + \frac{1}{2} V c_{k_1\downarrow}^\dagger c_{d \downarrow} \ket{\uparrow \downarrow} \bra{\uparrow \downarrow} V c_{d \uparrow}^\dagger c_{k_2 \uparrow} \Big(\frac{1}{\frac{B}{2} -\frac{U_d}{2}} + \frac{1}{-\frac{B}{2} -\frac{U_d}{2}} \Big) \Big]\\
	& = \sum_{k_1,K_2} \Big[J_1 S_d\cdot S_{k_1,k_2} + J_1 S_d^z c_{k_1\downarrow}^\dagger c_{k_2 \downarrow} +  J_1 S_d^z S_{k_1,k_2}^z \\ 
    &\hspace{1.5cm}+ J_2 S_d\cdot S_{k_1,k_2} + J_2 S_d^z c_{k_1\uparrow}^\dagger c_{k_2 \uparrow} +  J_2 S_d^z S_{k_1,k_2}^z \Big] ~,
\end{align*}
where $J_1 = - \frac{2V^2}{\frac{B}{2} -\frac{U_d}{2}}$ and $J_2 = - \frac{2V^2}{-\frac{B}{2} -\frac{U_d}{2}}$~, and we have used $ n_{d\uparrow} + n_{d\downarrow} = n_d$ and $ S_d^z = \frac{1}{2} (n_{d\uparrow} - n_{d\downarrow})$. Considering only those terms that contribute to the Kondo Hamiltonian $\sum_{k_1,K_2} \Big[(J_1 + J_2) S_d\cdot S_{k_1,k_2} \Big]$, we obtain the Kondo coupling $J=J_{1}+J_{2}$ as 
\begin{equation}
J_1 + J_2 = - \frac{2V^2}{\frac{B}{2} -\frac{U_d}{2}} - \frac{2V^2}{-\frac{B}{2} -\frac{U_d}{2}} = \frac{4V^2}{U_d - B } + \frac{4V^2}{U_d + B} = \frac{8U_d V^2}{U_d^2 - B^2} ~.
\end{equation}
This obtains the Kondo temperature $T_{k}$ as
\begin{equation}
	T_k = D_0 \sqrt{2J\rho} \hspace{1mm} e^{-\frac{1}{2J\rho}} = \frac{U_d}{2} \sqrt{\frac{16 U_d V^2 \rho}{U_d^2 - B^2}} \hspace{1mm} e^{-\frac{U_d^2 - B^2}{16 U_d V^2 \rho}}~,\nonumber
\end{equation}
where we have taken $D_0 = \frac{U_d}{2}$.
\section{Spectral Height of Impurity in up-spin configuration}
\label{L8S}
\begin{figure}[!ht]
    \centering
    \includegraphics[scale=0.31]{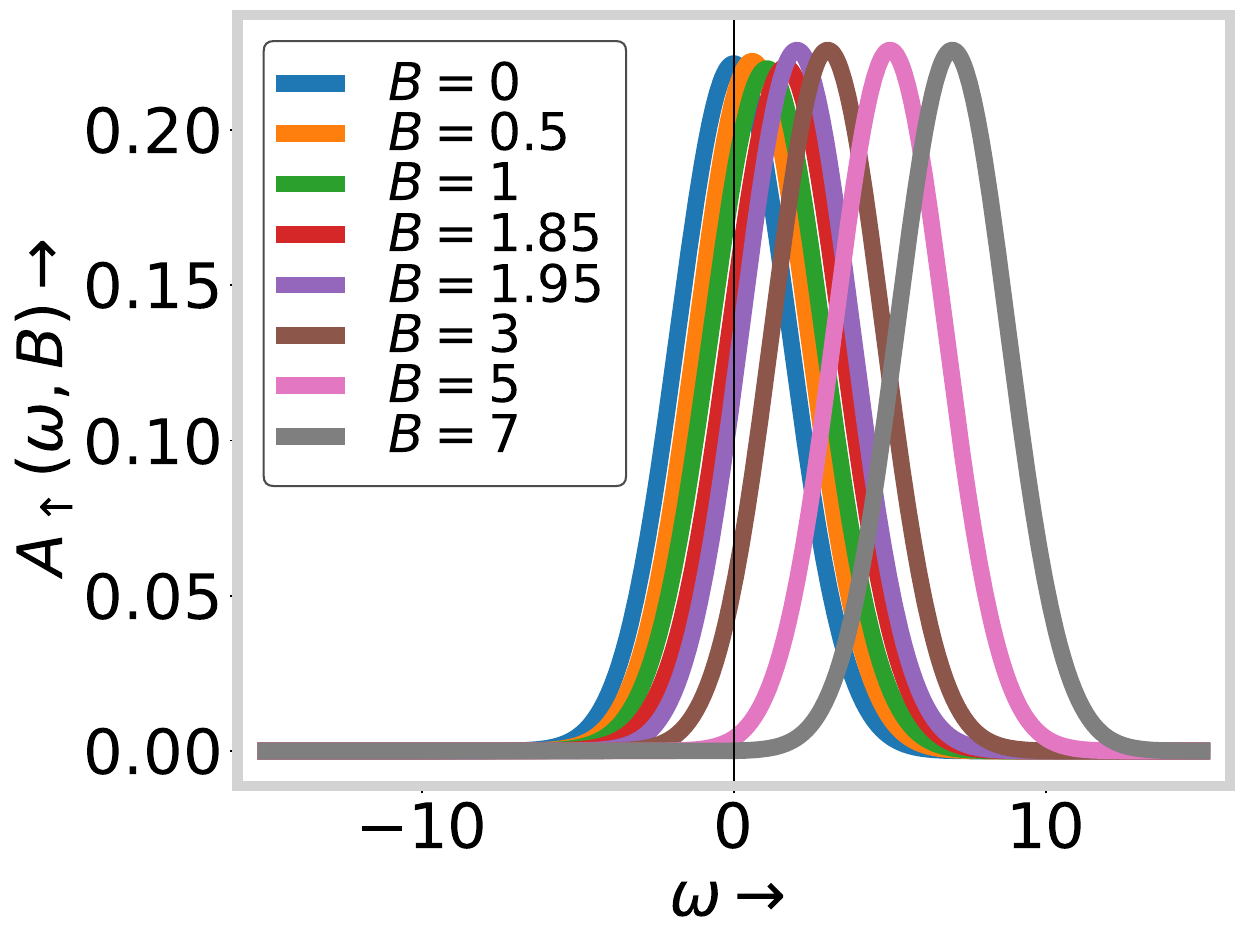}
    \includegraphics[scale=0.31]{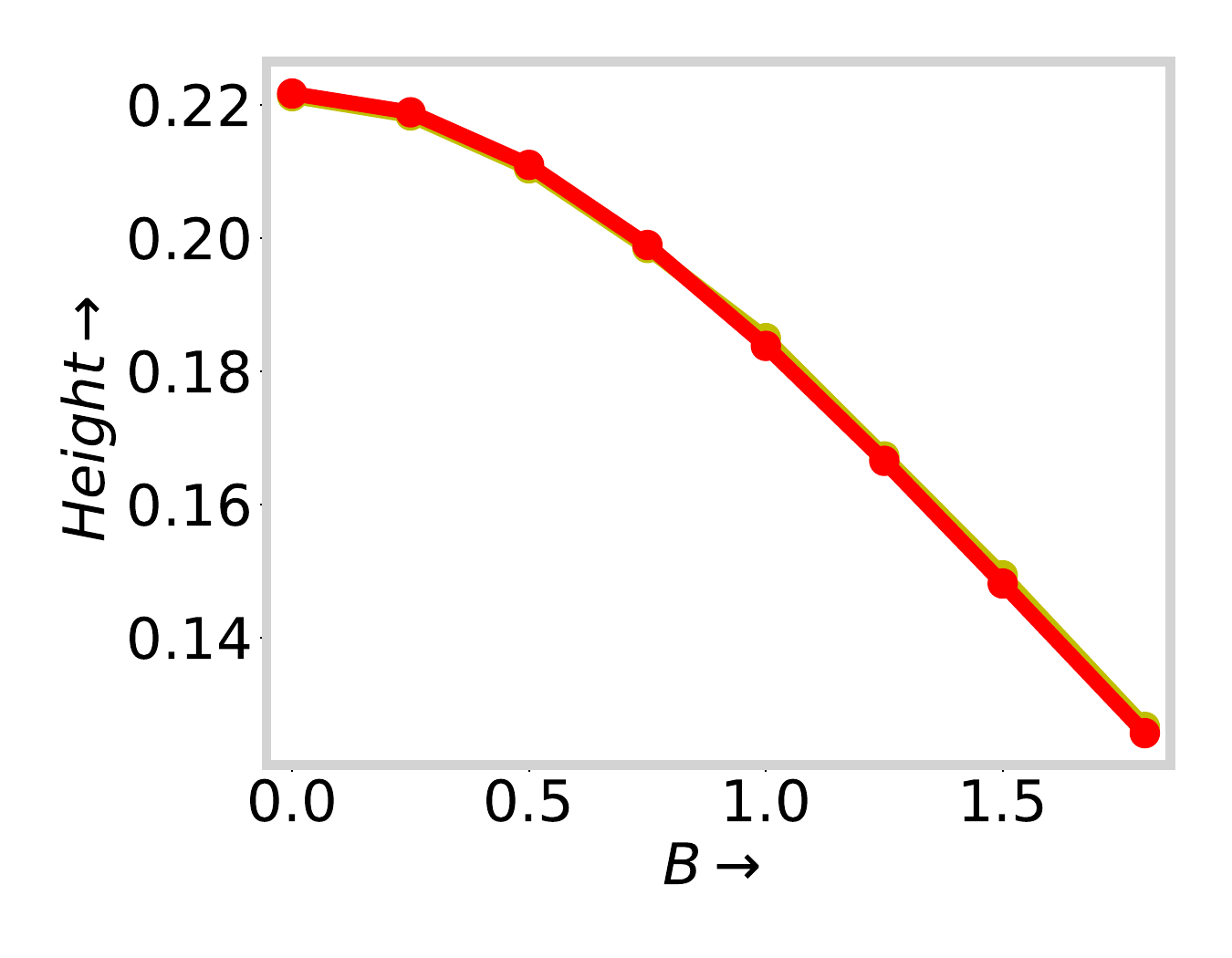}
    \caption{Variation of (left) up-spin impurity spectral function and (right) up-spin spertral height at $\omega = 0$ with local magnetic field.}
    \label{HYT}
\end{figure}
In Figure \ref{HYT} (Left), we plot the up-spin spectral function for the impurity as the local magnetic field is tuned. In Figure \ref{HYT} (Right), we plot the value of the up-spin spectral function at $\omega=0$, and the red line is a fit with an analytical expression~\cite{costi2000, andrei1982calculation} given by 
\begin{equation}
	A_{\sigma}(0,B) = \frac{1}{\pi^2 V \rho} \sin^2(\delta_\sigma(B))~,~ \delta_\sigma (B) = \frac{\pi}{2} -\alpha B + a \alpha^2 B^2~,
\end{equation}
where the parameters $\alpha =0.46$ and $a = 0.16$ with $V=0.608$ and $\rho =1$. It is interesting to note that the scattering phase shift $\delta_{\sigma}(B)$ is dependent on only a single dimensionful parameter $\alpha$.
\end{appendix}




\bibliography{KMmanuscript_Scipost.bib}

\nolinenumbers

\end{document}